\documentclass[prb,floatfix,aps,twocolumn,showpacs]{revtex4}
\usepackage{graphicx}
\usepackage[]{epsfig}
\usepackage[]{psfig}
\newcommand{\beq}{\begin{equation}}
\newcommand{\eeq}{\end{equation}}
\newcommand{\bea}{\begin{eqnarray}}
\newcommand{\eea}{\end{eqnarray}}
\newcommand{\nn}{\nonumber}

\newcommand{\eps}{\epsilon}

\newcommand{\D}{\Delta}

\newcommand{\dmi}{{1\over 2}}

\begin{document}

\title{Maximized Orbital and Spin Kondo effects in a 
single-electron transistor}

\author{Karyn Le Hur$^1$, Pascal Simon$^{2}$, and  L\'aszl\'o Borda$^{3,4}$}

\affiliation{$^1$ D\'epartement de Physique and RQMP,
 Universit\'e de Sherbrooke, Sherbrooke, Qu\'ebec, Canada, J1K 2R1}
\affiliation{$^2$ Laboratoire de Physique et Mod\'elisation des Milieux Condens\'es et \\
Laboratoire d'Etude des Propri\'et\'es Electroniques des Solides, CNRS\\
 25 av. des martyrs, 38042 Grenoble, 
France}
\affiliation{$^3$ Sektion Physik and Center for Nanoscience, LMU M\"unchen,
80333 M\"unchen, Theresienstr. 37}
\affiliation{$^4$ Research Group of the Hungarian Academy of Sciences, 
Institute of Physics, TU Budapest, H-1521}

\date{\today} 
\begin{abstract}
We investigate the charge fluctuations of a single-electron box 
(metallic grain)
coupled to a lead via a smaller quantum dot in the Kondo regime. The 
most interesting aspect of this problem resides in 
the interplay between {\it spin} Kondo physics stemming from the
screening of the spin of the small dot and {\it orbital} Kondo physics emerging
when charging states of the grain with (charge) $Q=0$ and $Q=e$ are almost 
degenerate. Combining Wilson's
numerical renormalization-group method with perturbative scaling approaches
we push forward our previous work [K. Le Hur and P. Simon, Phys. Rev. B
67, 201308R (2003)]. We 
 emphasize that for symmetric and slightly asymmetric 
barriers, the strong entanglement of charge and spin flip events in this setup 
inevitably results in a non trivial stable SU(4) Kondo fixed point near the 
degeneracy points of the grain. By analogy with a small dot sandwiched 
between two leads, the ground state is Fermi-liquid like 
which considerably smears out the Coulomb staircase behavior and 
hampers the Matveev logarithmic singularity to arise.
Most notably, the associated Kondo temperature $T_K^{SU(4)}$ might be
raised compared to that in the conductance experiments through a small
quantum dot $(\sim 1K)$ which 
makes the observation of our predictions {\it a priori} accessible. 
We discuss the robustness of the
SU(4) correlated state against the inclusion of an external magnetic field,
a deviation from the degeneracy points, particle-hole symmetry in
the small dot, asymmetric tunnel junctions and comment on the different 
crossovers.

\end{abstract}

\maketitle

\section{Introduction}

Recently, quantum dots have attracted a considerable interest due to their 
potential applicability as single electron transistors or as basic building 
blocks (qubits) in the fabrication of quantum computers.\cite{Loss} In the 
last years, a great amount of work has also been devoted
to studying the Kondo effect in mesoscopic structures.\cite{kouwenhoven}
A motivation for these
efforts was the recent experimental observation of the Kondo effect in
tunneling through a small quantum dot in the Kondo regime.\cite{Gold,Cron,vdw}
In these experiments, the excess electronic {\it spin} of the dot acts as a 
magnetic impurity. Let us also mention
that the manipulation of magnetic cobalt atoms on a copper surface and more
specifically the 
observation of the associated Kondo resonance via spectroscopy tunneling 
measurements\cite{Eigler,Crom} also stands for a remarkable 
opportunity to probe spin Kondo physics at the mesoscopic scale but in
another realm (not with artificial structures). 

A different set of problems relating the Kondo effect to the physics of
quantum dots is encountered when investigating the charge fluctuations 
of a large Coulomb-blockaded quantum dot (metallic grain).\cite{Aleiner} 
More precisely, one of the most 
important features of a quantum dot 
is the Coulomb blockade phenomenon, i.e., as a result of the
strong repulsion between electrons, the charge of a quantum dot
is quantized in units of the elementary charge $e$. Even a metallic dot at a 
micronmetric scale can still behave as a good 
single-electron transistor. When the gate voltage $V_g$ is increased the
charge of the grain 
changes in a step-like manner. This behavior is referred
to as a Coulomb staircase. 
Moreover when the metallic dot is weakly-coupled to a bulk lead, so 
that electrons
can hop from the lead to the dot and back, the dot charge remains to a large 
extent quantized. This quantization has been investigated with thoroughness
both theoretically\cite{Matv1,Grabert,Schon1,Schiller} and 
experimentally.\cite{Devoret}
It is important to bear in mind that such a problem is intrinsically connected
to an {\it orbital} or charge 
Kondo effect.\cite{Matv1} Indeed, near the degeneracy points of
the average charge in the grain one can effectively map the problem of 
charge fluctuations  onto a 
(planar)
two-channel Kondo Hamiltonian\cite{Cox,no,emery} with the two charge 
configurations in
the box playing the role of the impurity spin\cite{Matv1,Georg} and the 
physical spin of the
conduction electrons acting as a passive channel index. 
(This mapping is {\em a priori} 
valid only for weak tunneling junctions between the grain and the lead). For accessible
temperatures -- in general, larger than the level spacing of the
grain -- spin Kondo 
physics is not relevant.\cite{GHL} The quantity of interest 
is the average dot charge as a function of the voltage applied to a back-gate.
 Note that the average dot charge can be measured
with sensitivity well below a single charge.\cite{Konrad} Unfortunately, only
some fingerprints
of the two-channel Kondo effect were recently observed for a setting in 
semiconductor quantum dots.\cite{Berman} Indeed, the non-Fermi liquid nature 
of the two-channel Kondo effect
is hardly accessible in the Matveev's setup built on
semiconducting devices.\cite{Zarand3} On the one hand, the charging energy 
of the grain must
be large enough to maximize the Kondo temperature $T_K$ on the other hand
the level spacing must be small enough compared to $T_K$. It is difficult to
satisfy these two conflicting limits. A better chance
for observing the two-channel Kondo behavior may be reached 
if tunneling
between the lead and the grain involves a 
resonant level since it offers the possibility to actually 
enhance the Kondo temperature of the system.\cite{Eran}

\begin{figure}[ht]
\centerline{\epsfig{file=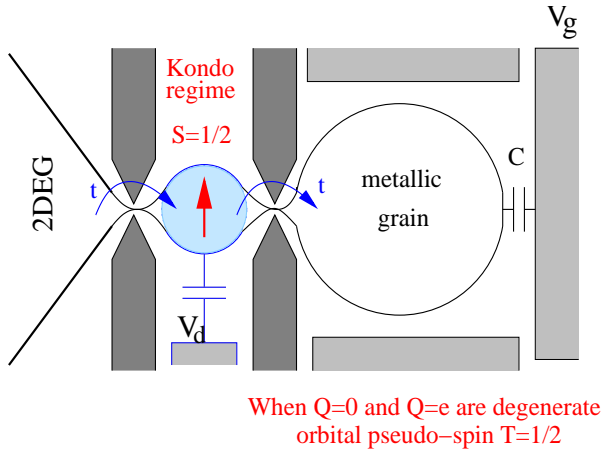,angle=0.0,height=5.9cm,width=
8.4cm}}
\caption{Schematic view of the setup. A micron scale grain (or large dot)  
is weakly coupled to a bulk
lead via a small dot in the Kondo regime which acts as an S=1/2 spin impurity.
The charges of the grain and the small dot are controlled by the gate voltage 
$V_g$ and $V_d$ respectively.
The auxiliary voltages can be used to adjust the tunnel junctions.}
\end{figure}

In this paper, the setup we analyze consists of a single-electron box or grain 
coupled to a reservoir through a smaller dot (Figure 1). We assume that the 
smaller dot contains an odd number of electrons and eventually acts 
as an S=1/2 Kondo impurity.\cite{kouwenhoven} Typically, when only charge
Kondo flips are involved the low energy physics near the degeneracy points
is well described by  a two-channel Kondo model,
in particular  the capacitance peaks of the grain
exhibit  at zero temperature a logarithmic singularity at the 
degeneracy points 
which ensures a nice Coulomb staircase even
for not too weak couplings between the quantum box and the lead.\cite{Matv1} 
In our setup, the Kondo effect has now two possible origins:
the spin due to the presence of the small dot playing the role of an S=1/2
spin impurity and the orbital degeneracy on the grain. Combining Wilson's
numerical renormalization-group (NRG) 
method with perturbative scaling approaches
we extend our previous work,\cite{su4} and emphasize that at (and near) the
degeneracy points of the grain, the two Kondo effects can be intertwined.
The orbital
degrees of freedom of the grain become strongly entangled with the spin degrees
of freedom of the small dot resulting in a stable fixed point with an SU(4)
symmetry. This requires symmetric or slightly asymmetric tunneling junctions. 
Furthermore, the low energy fixed point  is a Fermi-liquid  
which considerably smears out the Coulomb staircase behavior and 
hampers the Matveev logarithmic singularity to arise.\cite{Matv1} 
Remember that the major consequence of this enlarged symmetry in our
setup is that the grain capacitance exhibits instead of a logarithmic 
singularity, a strongly reduced peak as a function of the back-gate voltage, 
smearing charging effects in the grain considerably.
It is also worth  noting that the Kondo effect is 
maximized when both Kondo effects occur simultaneously. In particular, the
associated Kondo temperature $T_K^{SU(4)}$ can be strongly 
enhanced compared to that of the Matveev's original setup which 
may guarantee the verification of our predictions. We stress that the
Coulomb staircase behavior 
becomes  smeared out already in the weak tunneling limit due to the 
appearance
of spin-flip assisted tunneling. 
A different limit where
the small dot rather acts as a resonant level close to the Fermi level 
has been studied in Refs. [\onlinecite{Eran,Gramespacher}] where in contrast 
it was shown 
that the resonant level
has only a slight influence on the smearing of the Coulomb blockade 
even if the transmission coefficient through the impurity is one at resonance. 
This differs markedly from the case of an energy-independent 
transmission coefficient where the Coulomb staircase is completely destroyed
for perfect transmission.\cite{Nazarov,Matv1}
Furthermore, the charge of the grain in such a device can be used to measure 
the occupation of the dot.\cite{Gramespacher} The resonant-level 
behavior of Ref. [\onlinecite{Gramespacher}] is also recovered in our setup
when an orbital magnetic field is applied.

Let us mention that the possibility of a strongly correlated 
Kondo ground state possessing an SU(4) symmetry has also been discussed very
recently in the 
different context of two small dots coupled with a strong capacitive
inter-dot coupling.\cite{Zarand} The possibility of orbital and spin 
Kondo effects
in such a geometry was previously anticipated by 
Sch\"on {\it et al.}\cite{Schon}
inspired by preliminary 
experiments of Ref.~[\onlinecite{Weiss}]. It is worth noting
that these types of problems have also potential connections 
with the twofold orbitally degenerate 
Anderson impurity model\cite{Natan,Fabrizio} and more precisely
with the physics of certain 
heavy fermions like UBe$_{13}$ where the U ion is modeled
by a non-magnetic quadrupolar doublet\cite{Cox2} and then
quadrupolar (orbital) and spin Kondo effects can in principle 
interfere.\cite{Natan}

Our paper is structured as follows: In Section II, we resort to a Schrieffer-Wolff
transformation and derive the effective model including 
the different useful parameters. In Section III, assuming that we are far
from the degeneracy points of the grain we use a pedestrian perturbation 
theory; This reveals the importance of spin flips even in this limit. In 
Section IV, we carefully investigate both theoretically and numerically 
the interplay between orbital and spin Kondo effects at the degeneracy points.
In Section V, we discuss in details the effects of possible symmetry breaking 
perturbations and the crossovers generated by such perturbations.
Finally, section VI is devoted to the discussion of our results and especially
we summarize our main experimental predictions for such a setup.

\section{Model and Schrieffer-Wolff transformation}

In the sequel, we analyze in details the behavior of charge fluctuations in 
the grain. In order to model the setup depicted in Figure 1, 
we consider the Anderson-like Hamiltonian:
\begin{eqnarray}
\label{anderson}
H&=&\sum_k \epsilon_k a^\dag_{k\sigma} a_{k\sigma}+
\sum_p \epsilon_p a^\dag_{p\sigma} a_{p\sigma} + 
{\hat Q^2\over 2C}+\varphi \hat Q \nonumber
\\ 
&+& \sum_{\sigma} \epsilon a_{\sigma}^{\dag}a_{\sigma}+U 
n_{\uparrow}n_{\downarrow}
\\ \nonumber
&+& t\sum_{k\sigma}\left(a^\dag_{k\sigma}a_{\sigma}+h.c.\right)+
    t\sum_{p\sigma}\hbox{\huge{(}}a^\dag_{p\sigma}a_{\sigma}+h.c.
\hbox{\huge{)}},
\end{eqnarray}
where $a_{k\sigma}$, $a_{\sigma}$, $a_{p\sigma}$ are the annihilation
operators for electrons of spin $\sigma$ in the lead, the small dot, and
the grain, respectively, and $t$ is the tunneling matrix element 
which we assume  to be $k$ independent  for simplification. Let us first 
consider
 that tunnel junctions are {\it symmetric}. 
We  also assume that junctions are narrow enough and contain 
one transverse channel only. Extensions of the model to asymmetric or larger 
junctions 
will be analyzed later in section V.
We also assume that the energy spectrum in
the grain is continuous, which implies that the grain is
large enough such that its level
spacing $\Delta_g$ is very small compared to its charging energy 
$E_c=e^2/(2C)$:
$\Delta_g/E_c\rightarrow 0$ (in Ref.~[\onlinecite{Berman}] 
$\Delta_g\sim$ 70mK was not sufficiently small compared to the Kondo 
temperature scale
which hindered the logarithmic capacitance peak\cite{Matv1} to 
 completely develop).
$\hat Q$ denotes the charge operator of the grain, 
$C$ is the capacitance between the grain and the gate electrode, and $\varphi$
is related to the back-gate voltage $V_g$ through $\varphi=-V_g$. 
$\epsilon<0$ and $U$ are respectively
the energy level and charging energy of the small dot, and $n_{\sigma}=
a_{\sigma}^{\dag}a_{\sigma}$. The inter-dot capacitive coupling is assumed 
to be weak and therefore neglected.

We mainly focus on the particularly interesting situation where the small dot 
is 
in the {\it Kondo regime}, which requires  the last level to be singly 
occupied and the condition
\begin{equation}\label{Kondo}
t\ll -\epsilon,U+\epsilon,
\end{equation}
to be satisfied $(\epsilon<0)$. The resonant level limit 
where $\epsilon$ lies near the Fermi level will be addressed at some 
points in Section V. In the local moment regime, we can integrate out charge fluctuations in the small dot using a generalized Schrieffer-Wolff 
transformation.\cite{Schrieffer,Hewson}
More precisely, 
the system is described by the Hamiltonian:
\begin{eqnarray}
\label{ham}
H&=&\sum_{k} \epsilon_k a^\dag_{k} a_{k}+
\sum_p \epsilon_{p} a^\dag_{p}a_{p} + 
{\hat Q^2\over 2C}+\varphi \hat Q\\ \nonumber
&+&\sum\limits_{m,n} \left(
{J\over 2}\vec S\cdot{\vec{\sigma}}+V\right) a_{m}^\dag 
a_{n}.
\end{eqnarray}
To simplify the notations, the spin indices have been omitted and
hereafter. $m,n$ take values in the two
sets ``lead'' (k) or ``grain'' (p), the spin $\vec S$ is the spin 
of the small dot, $\vec{\sigma}$ are Pauli matrices acting on the spin space
of the electrons. Let us now discuss the parameters $J$ and $V$ in 
more details.

In the vicinity of one degeneracy point obtained for $\varphi=-e/2C$,
 where the grain
charging states with $Q=0$ and $Q=e$ are degenerate, we find explicitly:
\begin{equation}
J=2t^2\left[\frac{1}{-\epsilon}+\frac{1}{U+\epsilon}\right].
\end{equation}
A small direct hopping term 
\begin{equation}
V=\frac{t^2}{2}\left[\frac{1}{-\epsilon}-\frac{1}{U+\epsilon}\right],
\end{equation}
is also present and should not be neglected. In particular, this embodies
the so-called ``charge flips'' from the reservoir to the grain and vice-versa
in the Matveev's original problem. Notice that the
ratio $V/J$ can take values between -1/4 (when $U=-\eps$) and 
1/4 (when $U\to \infty$). $V=0$ corresponds to the particle-hole 
symmetric case where $2\epsilon+U=0$. For $\varphi=-e/2C$, the energy to add
a hole or an electron onto the metallic grain vanishes and therefore 
the Schrieffer-Wolff parameters $V$ and $J$ are completely identical to those 
of a small dot connected to  two 
metallic reservoirs.\cite{Raikh} Furthermore, remember that 
in the present model the ultraviolet 
cutoff at which the effective model becomes valid can be roughly
identified to $D\sim \hbox{min}\{E_c,\Delta_d\}$ where $\Delta_d$ is the
level spacing of the small dot (with today's technology it is possible to 
reach\cite{vdw} $\Delta_d\sim\ $2-3K and for the grain\cite{Berman}
$E_c\sim\ $2.3K).

On the other hand, far from the degeneracy point $\varphi=-e/2C$ 
-- which means on a charge plateau -- the energy to add a {\it hole} 
on the grain 
is $U_{-1}=E_c(1+2N)$ where $N=C V_g/e\neq 1/2$.
Similarly, it costs $U_1=E_c(1-2N)$ to add an extra {\it electron} onto the 
grain.
The lead-dot and grain-dot Kondo couplings, $J_0$ and $J_1$ respectively, 
then become {\it asymmetric} even for symmetric junctions:
\begin{eqnarray}\label{couplingss}
J_0 &=& 2t^2\left[\frac{1}{-\epsilon}+\frac{1}{U+\epsilon}\right]=J\\ \nonumber
J_1 &=& 2t^2\left[\frac{1}{U_1-\epsilon}+\frac{1}{U+\epsilon+U_{-1}}\right].
\end{eqnarray}
In the second equation, the virtual intermediate state where
  an electron first
hops from the grain onto the small dot induces
an excess of energy $U_{-1}$ in the second term. The first term contains
the energy of the intermediate state of the process where the temporal order 
of 
the hopping events is reversed. The off-diagonal 
terms where an electron from the reservoir [grain] 
flips the impurity spin and then jumps
onto the grain [reservoir] reads
\begin{eqnarray}
J_{01} &=& 2t^2\left[\frac{1}{U_1-\epsilon}+\frac{1}{U+\epsilon}\right]\\ 
\nonumber
J_{10} &=& 2t^2\left[\frac{1}{-\epsilon}+\frac{1}{U+\epsilon+U_{-1}}\right].
\end{eqnarray}
Note that in general particle-hole symmetry is absent in the large dot, so
in principle $J_{01}\neq J_{10}$. But, in our setting, 
$E_c=e^2/2C\ll |\epsilon|,U+\epsilon$, so in the following we will neglect
the asymmetry between $J_{01}$ and $J_{10}$ far from the degeneracy points 
$(J_{01}=J_{10})$ (this has no drastic consequence on the results) . In the finite temperature range $T< U_{1},U_{-1}$, these
off-diagonal processes are suppressed exponentially as 
$\tilde J_{10}=J_{10}(T)\approx J_{10}e^{-U_1/4kT}$, whereas the diagonal spin processes can be strongly renormalized at low temperatures. In other words, in
the Renormalization Group language, if we start at high temperature with a
set of Kondo
couplings $J_0, J_1, J_{01}, J_{10}$, the growing of $J_{01}, J_{10}$ is
cut-off when $T$ is decreased below $\hbox{max}(U_1,U_{-1})$ 
whereas the growing
of $J_0,J_1$ is not.
This offers a room to reach a 2-channel Kondo effect
in the spin sector (for asymmetric tunneling junctions) 
provided the condition $J_0=J_1$ can be reached with a fine-tuning of the 
gate voltages.\cite{oreg}
We can do the same approximation
for the $V$ term and define $V_{10}$, $V_{01}$ accordingly 
(with $V_{10}=V_{01}$) and also $\tilde V_{10}$.

\section{Pedestrian Perturbation Theory on a plateau}

We want first to compute the corrections to the
average charge on the grain on a charge plateau 
due to the Kondo and $V$ couplings 
bearing in mind that when the tunneling amplitude 
$t\rightarrow 0$, the average grain
charge $\langle\hat{Q}\rangle$ exhibits  perfect Coulomb staircase behavior 
as a function of $V_g$.
We confine ourselves to values of $\varphi$ in the range 
$-e/(2C)<\varphi<e/(2C)$,
which corresponds to the unperturbed (charge) value $Q=0$. 
A first natural approach is to assume that the Kondo and charge-flip 
couplings
are very small compared to the charging energy $E_c=e^2/(2C)$ of the grain
and to calculate the corrections to $Q=0$ in perturbation theory.
Despite this perturbative calculation will appear of limited use, 
it is very 
instructive to perform it in order to point the different sources 
of divergences
that appear when approaching  the degeneracy points, the main issue treated 
in this paper. At second
order, we find
\begin{equation}\label{pert}
\langle\hat{Q}\rangle_2\ =e \left({3\over 8} {J_{10}}^2+2{V_{10}}^2\right)
\ln\left({e/2C-
\varphi\over
e/2C+\varphi}\right).
\end{equation}
Note that at finite low temperature $T<U_1,U_{-1}$, we should use the 
renormalized off-diagonal couplings $\tilde J_{10},\tilde V_{10}$ which are 
small (in other words the flow of the off-diagonal Kondo couplings has been 
cut-off for 
$T<U_1,U_{-1}$). This better reproduces the (exact) numerical calculations of 
Ref.~\onlinecite{Schiller}.
For more details, we refer the reader to Appendix A. 
The density of states in the lead and in the grain have been assumed to be 
equal\cite{Matv1} and taken to be 1 for simplicity. 
This result tends to trivially generalize 
that of a grain directly coupled to a lead.\cite{Matv1}
However, there are two reasons that may suggest this perturbative approach is 
divergent. Higher-order terms -- already at cubic order--
involve logarithmic divergences
associated to the renormalizations of the 
Kondo couplings
(see Appendix A), but also other logarithms
indicating the vicinity of the degeneracy point $\varphi=- e/2C$
in the charge sector.
For example, a correction at cubic order to the result in Eq. (\ref{pert}) is 
given by 
\begin{equation}
\langle\hat{Q}\rangle_3 \propto{J_{0}{J_{10}}^2}\ln 
\left(\frac{D}{k_B T}\right)
\ln\left({e/2C-\varphi\over e/2C+\varphi}\right),
\end{equation}
We also have a similar correction in $J_1{J_{10}}^2$.
It would be potentially 
interesting to observe the logarithmic
temperature-dependence of $\langle\hat{Q}\rangle$ on a given plateau due to
Kondo spin-flip events. Note also that the perturbation theory in the $V_{10}$
term has been previously extended to the fourth order.\cite{Grabert}
The perturbative result is valid only far from the 
degeneracy points provided the renormalization, e.g., 
of the spin Kondo coupling $J_{0}$ is also cut-off either by 
the temperature $T$ or by a 
magnetic field $B$ [in general, for symmetric junctions one already gets
$J_0>J_1$ at the bare level; See Eq. (6)]. 
This considerably restricts the range of application of this 
perturbative calculation compared for example to the simpler setup 
involving a grain coupled to
a reservoir and even on a charge plateau 
the temperature much be larger than the emerging spin 
Kondo energy scale between the lead and the small dot. Finally, note that 
in our perturbative treatment at finite temperature $T<U_1,U_{-1}$
we have made the (standard) 
approximation: We
have only virtually introduced the temperature  
through the renormalization of the couplings $J_{10}$ and $V_{10}$.  

{\it The other regimes which requires non-perturbative 
approaches will be studied in Sections IV and V.}

\section{Orbital and Spin Kondo effects close to the degeneracy points}

In this section, we will be primarily interested in the situation close to 
the degeneracy 
point $\varphi=- e/2C$ 
where none of the perturbative arguments above can be applied. 
We want to show that the Hamiltonian given by Eq. (\ref{ham}) can be mapped 
onto some 
generalized Kondo Hamiltonian following Ref. [\onlinecite{Matv1}]. 

\subsection{Mapping to a generalized Kondo model}

Close to the degeneracy point $\varphi=- e/2C$ and for $k_B T\ll E_c$, only
the states with $Q=0$ and $Q=e$ are accessible and higher energy states 
can be removed from our theory introducing 
the projectors $\hat P_0$ and $\hat P_1$ (which project on the states 
with $Q=0$ and
$Q=e$ in the grain respectively). The 
truncated Hamiltonian (\ref{ham}) then reads:
\begin{eqnarray}
H&=&\sum\limits_{k,\tau=0,1} \epsilon_k 
a_{k\tau}^\dag a_{k\tau}\left(\hat P_0+\hat P_1\right)+eh\hat P_1\\ 
\nonumber
&+& \sum\limits_{k,k'}\hbox{\Huge{[}}\left({J\over 2}{\vec \sigma}\cdot\vec S
+V\right)\left(a_{k1}^\dag a_{k'0}\hat P_0+a_{k'0}^\dag a_{k1}\hat P_1\right)
\\ \nonumber
&+&\sum_{\tau=0,1} \left({J\over 2}{\vec\sigma}\cdot\vec
S+V\right)a_{k\tau}^\dag a_{k'\tau}\hbox{\Huge{]}},
\end{eqnarray}
where now the index 
$\tau=0$ indicates the reservoir and $\tau=1$ indicates the grain.
We have also introduced the small parameter 
\begin{equation}
h=\frac{e}{2C}+\varphi=\frac{e}{2C}-V_g\ll \frac{e}{C},
\end{equation}
which measures deviations from the degeneracy point.
Considering $\tau$ as an abstract {\it orbital} index, the Hamiltonian can be
rewritten in a more convenient way by introducing another set of Pauli
matrices for the orbital sector:\cite{Matv1,Georg}
\begin{eqnarray}
H&=&\sum\limits_{k,\tau} \epsilon_k a_{k\tau}^\dag a_{k\tau}+ehT^z 
\\ \nonumber
&+& \sum\limits_{k,k'}\hbox{\Huge{[}}     
\sum\limits_{\tau,\tau'}\left({J\over 2}{\vec\sigma}\cdot\vec S+V\right)
\hbox{\Large{(}}\tau^x T^x+\tau^yT^y\hbox{\Large{)}}_{\tau,\tau'}
a_{k\tau}^\dag a_{k'\tau'}
\\ \nonumber
&+ &\hskip 0.7cm 
\sum_\tau\left({J\over 2}{\vec\sigma}\cdot\vec S+V\right)a_{k\tau}^\dag 
a_{k'\tau}\hbox{\Huge{]}}.
\end{eqnarray}
In this equation, the operators $(S,\sigma)$ act on 
spin and the $(T,\tau)$ act on the (charge) orbital
degrees of freedom. 

The key role of this mapping stems from the
fact that $\langle\hat{Q}\rangle$ can be
identified as (an orbital pseudo-spin)
\begin{equation}\label{charge}
\langle\hat{Q}\rangle = 
e \left({1\over 2} + \langle T^z \rangle\right).
\end{equation}
Then, we can introduce the extra (charge) 
state $|Q\rangle$ 
as an auxiliary label to the state $|\Phi\rangle$ of the grain.
In addition to introducing the label $|Q\rangle$ we make the replacement
\begin{eqnarray}
 a_{k1}^\dag a_{k'0}\hat P_0 &\longrightarrow&  a_{k1}^\dag a_{k'0}T^{+}
\\ \nonumber
a_{k0}^\dag a_{k'1}\hat P_1 &\longrightarrow& a_{k'0}^\dag a_{k1}T^{-}.
\end{eqnarray}
Notice that $T^{+}$ and $T^{-}$ are pseudo-spin ladder operators acting only
on the charge part $|Q\rangle$. More precisely, we have the correct 
identifications 
\begin{eqnarray}
T^{-}|Q=1\rangle &=& T^{-}|T^z=+1/2\rangle\ = |Q=0\rangle \\ \nonumber
T^{+}|Q=0\rangle &=& T^{+}|T^z=-1/2\rangle\ = |Q=1\rangle,
\end{eqnarray}
meaning that the charge on the single-electron box is adjusted whenever a 
tunneling process takes place.
Furthermore, since $T^{+}|Q=1\rangle\ =0$ and  $T^{-}|Q=0\rangle\ 
=0$ these operators ensure 
in the same way as the projection operators $\hat P_0$ and 
$\hat P_1$ that only transitions between states with $Q=0$ and $Q=1$ take 
place. This leads us to identify $\hat{P}_1+\hat{P}_0$ with the identity 
operator on the
space spanned by $|0\rangle$ and $|1\rangle$ and $\hat{P}_1-\hat{P}_0$ with $2T^z$.
We now introduce an additional pseudo-spin operator
via:
\begin{eqnarray}
a_{k1}^\dag a_{k'0} &=& \frac{1}{2} 
a_{k\tau}^\dag\tau^- a_{k'\tau'}\\ \nonumber
a_{k0}^\dag a_{k'1} &=& \frac{1}{2} a_{k\tau}^\dag{\tau^+}a_{k'\tau'},
\end{eqnarray}
where the matrices $\tau^{\pm}=\tau^x\pm i\tau^y$ are standard combinations
of Pauli matrices. Finally, the Coulomb term $h$ mimics a magnetic field 
acting on the orbital space. Therefore, the (quantum) grain capacitance 
$C_q=-\partial\langle\hat{Q}\rangle/\partial h$ is equivalent to the
local isospin susceptibility $\chi_T=-\partial\langle T^z \rangle/\partial h$
up to a factor $e$. For simplicity, we will substract the classical 
contribution $C$ which is $V_g$-independent.

{\it But obviously, to compute the latter, we have to determine 
the nature of the Kondo ground state exactly.}

Typically, when only ``charge flips'' are involved through the $V$ term, 
the model can be mapped onto a two-channel Kondo model (the two channels
correspond to the two spin states of an electron), and 
the capacitance always exhibits
a {\it logarithmic divergence} at zero temperature.\cite{Matv1}
Here, we have a combination of spin and charge flips. Can we then expect two 
distinct energy scales for the spin and orbital sectors? To answer this 
question, we perform a perturbative scaling analysis following that of a
related model in Ref. [\onlinecite{Zarand2}].
We first 
rewrite the interacting part of the Hamiltonian in  real space as: 
\begin{eqnarray}\label{heff}
H_K &=& {J\over 2}\vec{S}\cdot \left(\psi^{\dag}{\vec{\sigma}}\psi\right) 
\nonumber
\\ 
&+& {V_z\over 2}T^z \left(\psi^{\dag}
{\tau^z}\psi\right)
+{V_{\perp}\over 2}\left[T^+ \left(\psi^{\dag}{\tau^-}\psi\right) +h.c.\right]
\\ \nonumber
&+& Q_z T^z\vec{S}\cdot \left(\psi^{\dag}{\tau^z}{\vec{\sigma}}\psi\right)
+ Q_{\perp}\vec{S}\cdot \left[T^+(\psi^{\dag}{\tau^-}{\vec{\sigma}}\psi)
+h.c.\right],
\end{eqnarray}
where $\psi_{\tau\sigma}=\sum_k a_{k\tau\sigma}$.

A host of spin-exchange $\otimes$ isospin-exchange interactions 
are clearly generated; $J$ refers to pure
spin-flip processes involving the S=1/2 spin of the small dot, 
$V_{\perp}$ to pure charge flips which modify the grain charge, and 
$Q_{\perp}$ describes exotic spin-flip assisted tunneling. 

This Hamiltonian exhibits a structure which is very similar
to the one introduced in Ref. [\onlinecite{Zarand}] 
in order to study
a symmetrical double (small) quantum dot structure 
with strong capacitive coupling.\cite{Zarand} 
However, since the physical situation
that led us to this Hamiltonian here 
is very different from that of Ref. [\onlinecite{Zarand}], 
our bare values for the coupling parameters are also very 
different (for $ J\ll 1)$: 
\begin{equation}
\label{couplings}
V_{\perp}=V,\ V_z=0\ ,\ Q_z=0\ ,\ Q_{\perp}=J/4.
\end{equation}
We have ignored the  potential scattering $V\psi^{\dag}\psi$ which
does not renormalize. It is also relevant to note that this
model belongs to the general class of problems of two coupled Kondo 
impurities. However the coupling between impurities, namely $Q_{\perp}$, 
is far different from the more usual RKKY interaction.\cite{Affleck}

Again, bear in mind that here the operators $\hat{P}_{1,0}=(1\pm
2T^z)/2$ and $\hat{p}_{0,1}=(1\pm\tau^z)/2$ project out the grain state
with $Q=e$ and $Q=0$, and the reservoir/grain electron channels, 
respectively. The spin $\vec{S}$ corresponds to the spin of the small dot
in the Kondo regime and the index $\sigma$ is the spin state of an electron
in the reservoir or in the grain. 

Note that in the situation of Ref.~[\onlinecite{Zarand}], the operators 
$\hat{P}_{\pm}=(1\pm
2T^z)/2$ and $\hat{p}_{\pm}=(1\pm\tau^z)/2$ rather project out the small double
dot states $(n_+,n_-)=(1,0)$ and $(0,1)$, and the right/left $(+/-)$
lead channels, 
respectively. Additionally, the spin $\vec{S}$ is the spin (excess) impurity 
either on the left or the right dot and the index $\sigma$ 
denotes the spin state of electrons in the reservoirs. The corresponding bare
values in that case
would be rather of the form:
\begin{equation}
V_{\perp}=Q_{\perp}\ ,\ V_z\ ,\ Q_z=J.
\end{equation}

\subsection{Perturbative Renormalization Group analysis}

The low-energy Hamiltonian can be treated  using perturbative 
renormalization group (RG).
It is relevant to observe that no new interaction terms are generated
to second order as the bandwidth is reduced.
By integrating out conduction electrons with energy larger than a scale
$E\ll D$ ($\sim \hbox{min}\{E_c,\Delta_d\}$ being either 
the level spacing of the 
small dot or the charging energy of the grain, i.e., 
the ultraviolet cutoff), we 
obtain at second order 
the following RG equations for the five dimensionless 
coupling constants:
\begin{eqnarray} \label{RG}
\frac{d J}{dl} &=& J^2 + {Q_z}^2 +2{Q_{\perp}}^2 \nonumber \\
\frac{d V_z}{dl} &=& {V_{\perp}}^2 +3 {Q_{\perp}}^2  \nonumber \\
\frac{d V_{\perp}}{dl} &=& V_{\perp}V_z+3 Q_{\perp}Q_z \\ 
\frac{d Q_z}{dl} &=& 2JQ_z +2V_{\perp}Q_{\perp}  \nonumber\\
\frac{d Q_{\perp}}{dl} &=& 2JQ_{\perp} + V_zQ_{\perp}+ V_{\perp} Q_z,\nonumber
\end{eqnarray}
with $l=\ln[D/E]$ being the scaling variable, $E$ is the running 
bandwidth. This RG analysis  
is applicable only very close to the degeneracy point $\varphi=-e/(2C)$
where the effective Coulomb energy in the grain or $h$ vanishes and obviously
only when all coupling constants stay $\ll 1$.
Higher orders in the RG have been neglected. 

Although the equations (\ref{RG})
have no simple analytic solution, one can try to read off
the essential physics from numerical integration and the initial 
conditions (\ref{couplings}).

Let us first discuss the most obvious case of a particle-asymmetric level, 
with $V_{\perp}>0$ meaning (large) $U\gg -2\epsilon$. In this 
case, the numerical integration of the RG flow indicates that
even though we start with completely {\it asymmetric}
bare values of the coupling constants, all
couplings diverge at the same energy scale due to the presence
of the spin-flip assisted tunneling terms $Q_{\perp}$ and $Q_{z}$.
This energy scale that we can identify with a generalized Kondo temperature
is difficult to calculate analytically. However, we can approximate it
by the one of the completely symmetrical model  
\begin{equation}
T_K^{SU(4)}\sim D\ e^{-1/4J}. 
\end{equation}
Furthermore, we have checked numerically that 
all coupling ratios  converge to 
one in the low energy limit provided the RG equations can be 
extrapolated in this regime. These results have been summarized in Figure~2.
As confirmed below with an exact Numerical RG treatment,
the entanglement of spin and orbital degrees of freedom in this geometry
will lead to an higher symmetry than $SU(2)\otimes SU(2)$, namely SU(4), and
then to the formation of a Fermi-liquid correlated ground-state with, e.g.,
the complete
screening of the orbital spin $\vec{T}$. [SU(4) is the minimal group allowing
spin-orbital entanglement and which respects rotational invariance both
in spin and orbital spaces.]
Recall that the presence of the 
spin-flip assisted tunneling terms then definitely hinders the possibility of a
non-Fermi liquid ground state induced by the over-screening of the the 
pseudo-impurity $\vec{T}$.

\begin{figure}[ht]
\centerline{\epsfig{file=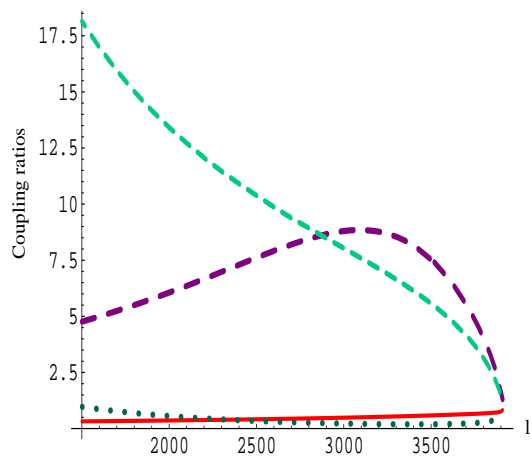,angle=0.0,height=6cm,width=
6.9cm}}
\caption{Evolution of the four coupling ratios as a function of the scaling
variable $l=\ln(D/E)$. The initial conditions have been chosen as: 
$J(0)=u$, $
V_{\perp}(0)=0.10u$, $Q_{\perp}(0)=u/4$ with $u=0.00018$ and 
$V_z(0)=Q_z(0)=0$. The full line
is $Q_{\perp}/J$, the dotted line $V_{\perp}/V_z$, and the dashed lines to
$Q_{\perp}/V_{\perp}$ and $Q_{\perp}/Q_z$ 
(which diverges for $l\rightarrow 0$). All the couplings are strongly
renormalized for 
$l_c\approx 3914$ and all their ratios converge to 1. Extrapolating the flow
to $l\gg l_c$ would give a straight horizontal line where  the coupling ratios 
remain $1$.} 
\end{figure}

Let us now  analyze the  particle-hole symmetric case, 
i.e., $V_{\perp}=0$. At second order, the RG flow would tend to
suggest that two parameters, namely $V_{\perp}$ and $Q_{z}$ remain
{\it zero} whatever the energy scale. Typically, the Kondo coupling $J$
is the largest throughout the RG flow and seems to be the first one to 
diverge. On
the other hand, the ratios $V_z/J$ and $Q_{\perp}/J$ cannot be neglected 
which tends to
exclude an $SU(2)\times SU(2)$ symmetry where the spin and the orbital
degrees of freedom would be independently screened (Figure 3). Instead, spin-orbital 
mixing (entanglement) seems to be prominent at low-energy.
Even though the perturbative RG is certainly
not sufficient to draw more definitive conclusions, it is 
also instructive to observe that for $V_{\perp}<0$, the ratios $Q_{\perp}/J$
and $V_z/J$
still converge to one. Since the system definitely has to restore
the rotational invariance both in spin and orbital spaces, this
tends to emphasize that higher-order
terms play a crucial role in the crossover regime eventually 
by restoring
an SU(4) Fermi-liquid even for those cases. Moreover, 
the RG analysis suggests
that the temperature 
scale at which the Fermi-liquid behavior  emerge would be much smaller for 
vanishing and negative $V_{\perp}$ because the system needs a much longer 
time to restore the rotational symmetry both in spin and orbital spaces. 
To
enumerate higher order terms would be a very tedious task, therefore 
this assertion 
will be rather checked by NRG, a completely non perturbative method. 
To summarize this part, we emphasize that 
for $V_{\perp}\leq 0$, the above perturbative analysis does not allow us
to determine  the  precise nature of the low-temperature
fixed point, whether the orbital (isospin) 
moment is exactly screened or over-screened.
We will prove in section \ref{secNRG} using NRG, that a  SU(4) strongly-correlated ground
state  emerges for any physical value of $V_{\perp}$, i.e., $-J/4\leq
V_{\perp}\leq J/4$. 

\begin{figure}
\centerline{\epsfig{file=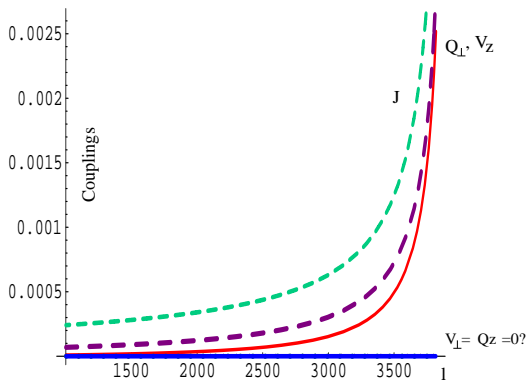,angle=0.0,height=5cm,width=
7cm}}
\caption{Here, we have chosen: 
$J(0)=u$, $Q_{\perp}(0)=u/4$ with $u=0.00018$ and 
$V_{\perp}(0)=V_z(0)=Q_z(0)=0$. The coupling $J(l)$ is the largest throughout
the RG flow, but the ratios $Q_{\perp}/J$ and $V_z/J$ cannot be neglected. 
Furthermore, at second order, the couplings $V_{\perp}$ and $Q_z$ would
remain zero. However, the NRG  concludes that even in this limit the system
converges to an SU(4) Fermi-liquid 
fixed point with identical coupling constants which emphasizes the importance
of higher-order terms and that
spin-orbital mixing is well prominent and that the rotational invariance is
restored both in spin and orbital spaces.}
\end{figure}

\subsection{Entanglement of spin and charge degrees of freedom}

This RG analysis suggests -- at least for not too small positive 
$V_{\perp}$ -- that
our model becomes equivalent at low energy to an SU(4) {\it symmetrical} 
exchange model:
\begin{eqnarray} \label{irrrep}
\hskip -1.2cm
H_K \hskip -0.15cm &=& \hskip -0.15cm 
J \sum\limits_{A} \psi^{\dag}_{\mu} t^A_{\mu\nu}
\left[\sum\limits_{\alpha\beta}\left(S^{\alpha}+{1\over 2}\right)
\left(T^{\beta} +{1\over 2}\right)\right]^A\psi_{\nu} \\ \nonumber
 \hskip -0.15cm &=&  \hskip -0.15cm {J\over 4}
\sum\limits_{A} M^A\sum\limits_{\mu,\nu} \psi^{\dag}_{\mu}
t^A_{\mu\nu}\psi_{\nu}.
\end{eqnarray}
Since all the coupling ratios converge to
one, we have rewritten the Kondo Hamiltonian (\ref{heff}) with 
the unique coupling constant $J$.
We have introduced the ``hyper-spin'' 
\begin{equation}
M^A\in \left\{2S^{\alpha},2T^{\alpha},
4S^{\alpha}T^{\beta}\right\},
\end{equation}
for $\alpha,\beta=x,y,z$. The operators $M^A$ can be regarded
as the 15 generators of the SU(4) group. 
Moreover, this conclusion will be strongly 
reinforced by the NRG analysis proposed below 
(whose range of validity is broader than Eqs. (\ref{RG})) which indeed
concludes that the effective Hamiltonian (\ref{irrrep}) is appropriate for 
all values of $-J/4\leq V_{\perp}\leq J/4$. Note that apparently (\ref{irrrep})
has $SU(2)\times SU(2)$ symmetry, representing rotational invariance in both
spin and orbital (pseudo-spin) spaces, and also interchange symmetry between
spin and pseudo-spin. But, the full symmetry is actually the higher symmetry
group SU(4), which clearly unifies (entangles) the spin of the small dot 
and the charge degrees of freedom of the metallic grain. 
Notice that the irreducible 
representation of SU(4) written in Eq. (\ref{irrrep}) has been used previously 
for spin systems with orbital 
degeneracy.\cite{Li,Azaria}
The electron operator $\psi$ now transforms under the fundamental 
representation of the SU(4) group, with generators $t^A_{\mu\nu}$  
$(A=1,...,15)$, and the index $\mu$ labels the four combinations of
possible spin $(\uparrow,\downarrow)$ 
and orbital indices $(0,1)$, which means $(0,\uparrow)$, $(0,\downarrow)$,
$(1,\uparrow)$ and $(1,\downarrow)$. 

The emergence of such a
strongly-correlated SU(4) ground 
state, characterized by the quenched hyper-spin operator
\begin{equation}
\left(\vec{S}+{1\over 2}\right)
 \left(\vec{\hbox{T}}+{1\over 2}\right),
\end{equation}
clearly reflects the 
strong {\it entanglement} 
between the {\it charge} degrees of freedom of
the {\it grain} and the {\it spin} degrees of freedom of the {\it small dot} 
at low energy induced by the prominence of spin-flip assisted tunneling. 
There
is the formation of an SU(4) Kondo singlet which is a singlet of the spin 
operator, 
the orbital operator,
and the orbital-spin mixing operator $U^{\alpha,\beta}=S^{\alpha}T^{\beta}$.
Again, let us argue that this enlarged symmetry arises whatever 
the parameter $V_{\perp}$ simply because the spin-flip assisted 
tunneling term 
$Q_{\perp}$ always flows off to strong couplings at the same time than the
more usual Kondo term $J$; The system then must inevitably converge to a fixed
point with orbital-spin mixing. To respect rotational 
invariance in both spin
and orbital spaces the only possibility is indeed an SU(4)-symmetric Kondo
model (as agreed with NRG).

\subsection{Capacitance: Destruction of Matveev's logarithmic singularity}

The (one-channel) SU(N) Kondo model has been extensively studied in the 
literature (see, e.g., Ref. [\onlinecite{Bickers}]). In particular, the 
strong coupling regime 
corresponds to a {\it dominant} Fermi liquid fixed point induced by the
complete screening of the hyper-spin $M^a$, implying that 
all the generators of SU(4) yield a local susceptibility with a behavior 
in\cite{Parcollet} $\sim 1/T_K^{SU(4)}$. $T^z$ being one of these 
generators, we deduce
 that 
$\chi_T=-\partial\langle T^z \rangle/\partial h$ and then the (quantum)
capacitance
of the grain ${C}_q=-\partial\langle\hat{Q}\rangle/\partial h$ roughly 
evolves as $1/T_K^{SU(4)}$ at low
temperatures;\cite{Parcollet} We have substracted the classical capacitance $C$. Consequently, for $h\ll e/C$,
we obtain a linear 
dependence of the average grain charge as a function
of $V_g=-\varphi$:
\begin{equation}\label{chargegrain}
\langle\hat{Q}\rangle -\frac{e}{2} =-e\frac{h}{T_K^{SU(4)}}
=-\frac{e}{T_K^{SU(4)}}\left[{e\over 2C}+\varphi\right].
\end{equation} 

The hallmark of
the formation of the SU(4) Fermi liquid 
in our setup is now clear. The (grain) capacitance
peaks are completely smeared out by the mixing of spin and charge flips and
 Matveev's logarithmic singularity\cite{Matv1} has been completely destroyed.
Additionally, the strong renormalization of the $V$ (and $J$) term 
-- and the stability of the strong-coupling Kondo fixed point -- clearly 
reflects that the effective transmission coefficient 
between the lead and the grain becomes maximal close to 
the Fermi level. 
(The maximum of the tunneling appears not exactly at the Fermi level
as one could guess from the value of the phase shifts $\delta=\pi/4$.). 

This example  could also be interpreted
as an interesting proof that one can already wash out the 
Coulomb staircase when the `effective' transmission coefficient between the 
grain and the lead is roughly one only close to
the Fermi energy 
(and not for all energies\cite{Nazarov}). Conceptually, this 
is not accessible
with a small dot in the resonant level limit.\cite{Eran,Gramespacher} 
{\it We stress that this stands for a remarkable signature of the formation of
a Fermi-liquid ground state when tunneling through a single-electron box.}

\subsection{Confirmation with Numerical Renormalization Group 
analysis}
\label{secNRG}

In order to confirm the results obtained by perturbative RG and extend our
investigation to the strong coupling regime, we have performed
a collaborative NRG\cite{NRG,Hewson} analysis of the model 
described by Eq.~(\ref{heff}) similar as in Ref.~[\onlinecite{Zarand}]. 
Note in passing that the model of Eq.~(\ref{heff}) 
with asymmetric bare values is not strictly speaking integrable. 
Therefore, we resort to the NRG method which in general
can be successfully applied to (various) 
two-impurity Kondo models.\cite{Jones2}
 At the heart of the NRG approach
is a logarithmic energy discretization of the conduction band around the Fermi
points. 
In this method -- after the logarithmic
discretization of the conduction band and a Lanczos transformation -- one 
defines a sequence of discretized Hamiltonians, $H_N$, with the 
relation:\cite{NRG}
\begin{equation}
{H}_{N +1} \equiv\Lambda^{1/2}{H}_N+\sum_{\tau\sigma}
\xi_{N}\left(f^{\dagger}_{N,\tau\sigma}f_{N+1,\tau\sigma}+ \mbox{h.c}
\right), 
\label{eq:rec}
\end{equation}
where $f_{0,\tau\sigma} = \psi_{\tau\sigma}/\sqrt{2}$ and 
$H_0 \equiv 2\Lambda^{1/2}/(1 + \Lambda)\; H_{\rm K}$ 
with $\Lambda\sim3$ as discretization parameter, and
$\xi_N\approx 1$. For the definition of $f_N$
see Ref.~[\onlinecite{NRG}]. The original Hamiltonian is connected to 
the $H_N$'s as
$
H = \lim_{N\to\infty} \omega_N H_N
$ 
with  $\omega_N = \Lambda^{-(N+1)/2}(1 + \Lambda)/2$. Using the 
logarithmic separation of the energy scales we are allowed to
diagonalize $H_N$'s iteratively and calculate physical quantities
directly at the energy scale $\omega\sim\omega_N$. 
We have calculated the dynamical spin and orbital spin (ac) susceptibilities 
\begin{equation}
{\Im m}\chi_{\cal O}(\omega)={\Im m}{\cal F}
\langle [{\cal O}(t),{\cal O}(0)]\rangle\;,
\label{eq:suscept}
\end{equation}
where ${\cal O}=T^z,S^z$ and $\cal F$ denotes the Fourier transform.
According to the discussion above, the couplings were chosen
as $J=4Q_\perp$, $Q_z=V_z=0$. 

\begin{figure}[ht]
\centerline{\epsfig{file=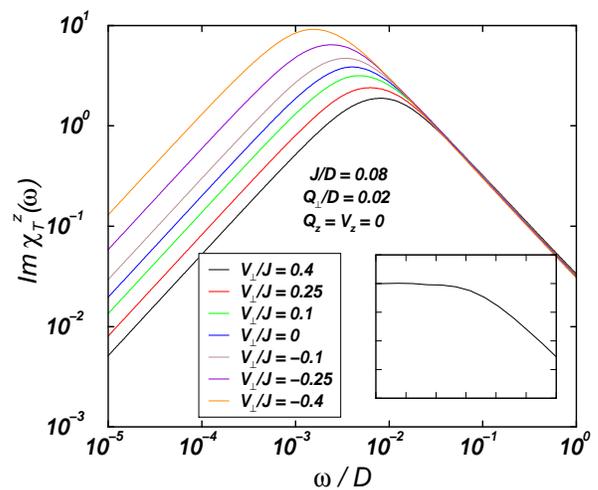,angle=0,height=6.7cm,width=7.8cm}}
\vskip -0.38cm
\caption{\label{fig:su4V}
The orbital spin $T^z$ susceptibility for different values of
$V_\perp$. In all the cases the susceptibility shows a typical
SU(4) 
Fermi-liquid state at $\omega=T_K^{SU(4)}(V_\perp)$. {\it Inset }: As a 
comparison we plot
the same quantity for the 2-channel Kondo model. Furthermore, we can clearly
observe that $T_K^{SU(4)}$ markedly decreases for lower values of $U$ 
meaning $V_{\perp}/J<0$ (i.e., by making the small dot larger 
and larger.\cite{karyn3})}
\end{figure}

The obtained orbital spin 
susceptibility for different values of $V_\perp$ is shown in 
Figure~\ref{fig:su4V}. 
Regardless of the value of $V_\perp$, the $T^z$ 
susceptibility exhibits a typical Fermi-liquid like peak at an energy scale 
which can be identified as $T_K^{SU(4)}$. 
Above this
energy scale it behaves as $\chi\sim\omega^{-1}$
indicating that the correlation function
in Eq.~(\ref{eq:suscept}) is constant for very short times 
while for $\omega < T_K^{SU(4)}$, $\chi\sim\omega$
as a signature of the $\sim 1/t^2$ asymptotic of the
afore mentioned correlation function for a Fermi-liquid model. Indeed, at T=0, 
this ensures a grain's capacitance,
\begin{equation}
{C}_q=\int_{1/T_K^{SU(4)}}^{+\infty} dt\
\langle [T^z(t),T^z(0)]\rangle=\frac{1}{T_K^{SU(4)}}\cdot
\end{equation}
Furthermore, as one can see in Figures~\ref{fig:su4V},
\ref{fig:chi_S_z_Delta_var} (for $\Delta_z\rightarrow 
0$) the Kondo screening simultaneously 
takes place in the 
spin and orbital sectors,
indicating the $SU(4)$-symmetric nature of the effective low energy
Hamiltonian. 
\hskip -0.1cm

\hskip -0.3cm
To give a rigorous proof of the $SU(4)$ Fermi liquid 
ground state one has to analyze the finite size spectrum obtained by
NRG. It turns out that (as in Ref.~[\onlinecite{Zarand}]) 
the spectrum can be understood as a sum of 
four independent chiral fermion spectra with phase shift $\pi/4$
in accordance with the prediction of the $SU(4)$ Fermi liquid 
theory.
This result proves that the low-energy behavior is described by 
the Fermi liquid theory even at $V_\perp=0$, but as conjectured above
the temperature scale at which the Fermi liquid emerges decreases
as we change the coupling $V_\perp$ from $0.4J$ to $-0.4J$.

For comparison, in the inset of Figure~\ref{fig:su4V} we plot
the dynamical susceptibility for the two channel Kondo model: In that case,
$\Im{m}\chi(\omega)\sim const.$ which in contrast traduces 
that the capacitance $C_q$ would exhibit a logarithmic divergence at zero 
temperature.\cite{Matv1} 

Additionally, the SU(4) Kondo temperature scale 
is considerably reduced for negative values of $V_{\perp}$, i.e., by 
decreasing the on-site interaction $U$ on the small dot $(U\ll -2\epsilon)$.
This makes sense 
since by substantially decreasing the Coulomb energy
of the small dot -- i.e., by progressively increasing the size of the small
dot -- one expects the breakdown of the SU(4)
fixed point and a situation similar to that of a reservoir and {\it two}
large dots\cite{karyn3} (According to Eq.~(\ref{Kondo}), spin
Kondo physics should definitely vanish for $U\ll -\epsilon$).

\section{On the stability of the SU(4) fixed point and crossovers}

In contrast to the two-channel Kondo fixed point, which is known to be 
extremely fragile with respect to perturbations (e.g., channel asymmetry,
magnetic field), the SU(4) fixed point is robust at least for {\it weak}
perturbations. 

In order to demonstrate the robustness of the SU(4) Fermi liquid 
fixed point 
we have checked the role, e.g., of a magnetic field in real and orbital spin
sectors. It turns out that both terms are marginal operators in RG sense.
On the other hand, when the magnetic [orbital] field is much 
larger than the scale of the 
Kondo temperature, the processes which involve spin [orbital spin] flips are
suppressed and low energy physics is described by
a {\it one-channel} orbital spin [spin] Kondo effect, with a smaller Kondo 
temperature than that of the SU(4) case. Let us now thoroughly analyze
the different fixed points, the effects of an asymmetry between the
tunnel junctions and of rather large junctions with more conducting channels. 

\subsection{Magnetic field}

First of all, we have checked with NRG that the SU(4) Fermi liquid fixed point
resists for quite weak external magnetic field.
But, applying a {\it strong} magnetic field $B\gg T_K$ unavoidably
destroys the SU(4) symmetry. But

\begin{figure}[ht]
\centerline{\epsfig{file=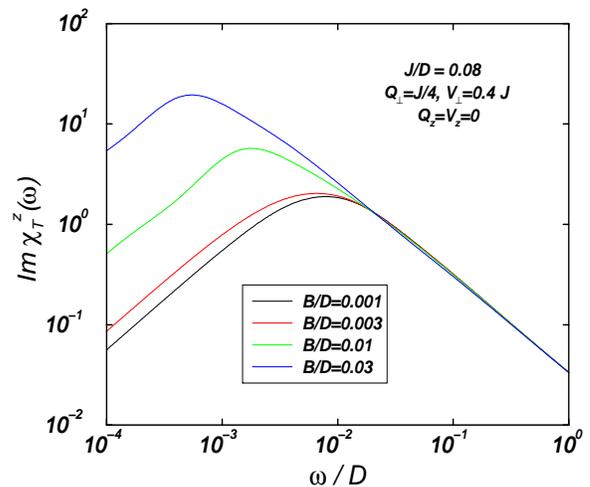,angle=0,height=6.7cm,width=7.7cm}
}
\caption{\label{fig:chi_T_z_B_var}
The orbital spin $T^z$ susceptibility for different values of the
external magnetic field $B$. The low-energy physics consists of a Fermi
liquid regardless of $B$, but the symmetry is reduced for large
magnetic fields to $SU(2)$ (for the orbital space) and the Kondo energy scale
as well.}
\end{figure}

\hskip -0.3cm at zero temperature, we expect 
the behavior of charge fluctuations close to the degeneracy points 
to remain 
qualitatively similar. 
Indeed, in a large magnetic field spin flips 
are suppressed at low temperatures, 
i.e., $Q_{\perp}=Q_z=J=0$, and the orbital degrees of freedom,
through $V_{\perp}$ and $V_z$, develop a standard one-channel Kondo model
(the electrons have only spin-up or spin down), which also results in 
a Fermi-liquid ground state with a linear dependence of the average grain 
charge as in Eq. (\ref{chargegrain}). Yet, the emerging Kondo temperature 
will be much smaller, 
\begin{equation}
T_K[B=\infty]\approx D \ e^{-1/V},
\end{equation}
with for instance $V\approx t^2/(-2\epsilon)$ for $U\rightarrow +\infty$,
and might not be detectable experimentally. A substantial
decrease of the Kondo temperature when applying an
external magnetic field $B$ has also been certified 
using NRG even for extremely large values of $V$ (Figure 5). 

\subsection{Away from the degeneracy points: small dot as a resonant level}

A weak orbital magnetic field (orbital splitting) $\Delta_z\propto h$ does not
modify the SU(4) Fermi liquid state. 

Moreover, the application of a {\it strong} $\Delta_z$ 
always leads to a single-channel Kondo effect in the spin 
sector. A naive consideration -- focusing on the RG flow above -- would 
suggest the possibility of a two-channel (spin) Kondo 
effect: The simultaneous screening of the excess 
spin of the small dot by the lead and the 
grain electrons, independently. However, going back to the Schrieffer-Wolff 
transformation for the situation away from the degeneracy points, the charging
energy of the metallic grain 

\begin{figure}[ht]
\centerline{
\epsfig{file=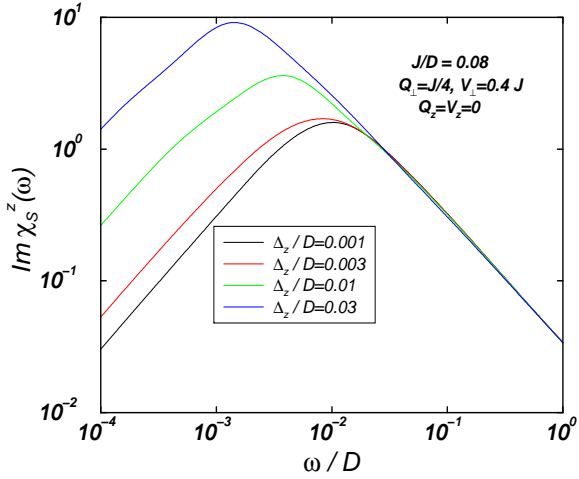,angle=0,height=6.6cm,width=7.7cm}
}\caption{\label{fig:chi_S_z_Delta_var}
The real spin $S^z$ susceptibility for different values of the
orbital splitting, $\Delta_z$. 
For $\Delta_z>T_K^{SU(4)}$ the processes which involve orbital spin flip are
suppressed resulting in a purely {\it one-channel} spin 
Kondo effect with a smaller Kondo 
temperature of the order of that for a small dot embedded between two 
leads $T_K[\Delta_z]$. Recall that the energy scale at
which the SU(4) correlated state arises can be much larger than 
$T_K[\Delta_z]$ 
which should certainly ensure the observation of our theoretical results. 
It is
worthwhile to note
the parallel between Figures 5 and 6 by interchanging $T^z\leftrightarrow 
S^z$ and $\Delta_z\leftrightarrow B$ (however, $T_K[\Delta_z]>T_K[B]$).}
\end{figure}

\begin{figure}[ht]
\centerline{
\epsfig{file=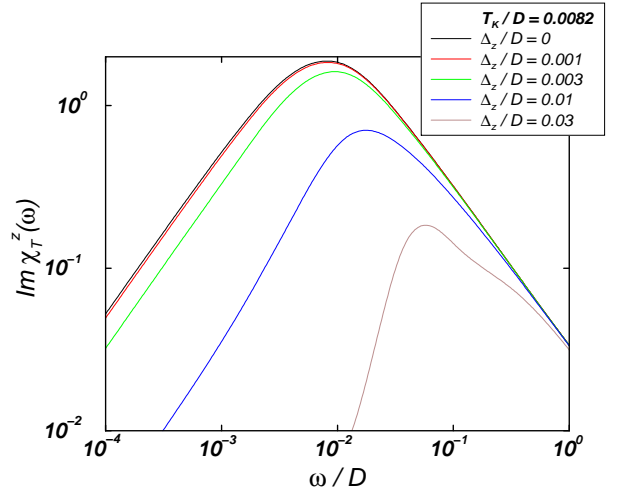,angle=0,height=6.9cm,width=8cm}
}
\caption{
The orbital spin $T^z$ susceptibility for different values of the
orbital splitting, $\Delta_z$. 
For $\Delta_z>T_K^{SU(4)}$ the processes which involve orbital spin flip are
clearly suppressed at a scale of $\Delta_z$ producing instead 
a Schottky anomaly. The orbital pseudospin model then becomes inappropriate to 
describe the charge fluctuations of the grain at low energy. We rather
apply another resonant level mapping and a 
perturbation theory similar to that of Ref.~[\onlinecite{Gramespacher}].}
\end{figure}

\hskip -0.3cm 
definitely ensures $J_1\neq J_0$ (provided we start with almost symmetric 
junctions), a condition that destroys the stability of the two-channel spin 
Kondo fixed point. The spin Kondo
coupling $J_0$ will be the first one to flow off to strong couplings 
(as anticipated in Section III).
The NRG calculation clearly 
confirms this expectation: the $\Delta_z$ term not only suppresses the
orbital spin-flip terms but also generates an asymmetry between the 
grain-dot and lead-dot spin couplings which destroys the two-channel
Kondo behavior. The possible two-channel (spin) Kondo regime
proposed by Oreg and Goldhaber-Gordon\cite{oreg} can not be reached with
this model, at least, for symmetric junctions. Asymmetric junctions and a fine
tuning of the grain gate voltage far from the degeneracy points would be 
necessary to reach the condition $J_0=J_1$.
 On the other hand we will see that, for quite asymmetric barriers,
a two-channel Kondo behavior rather for the orbital degrees of freedom 
can appear near the degeneracy points but at extremely small (and {\it a priori} unreachable) 
temperatures.

For $\Delta_z\gg T_K^{SU(4)}$, the Kondo temperature scale
here resembles that for a small dot connected to two 
leads $(J_0=J)$\cite{Raikh} and, in principle, is still experimentally 
accessible:
\begin{equation}
T_K[\Delta_z]\sim D\ e^{-1/J}<T_K^{SU(4)}.
\end{equation}
Henceforth, this will cutoff the 
logarithmic divergence in the charge fluctuations 
away from the degeneracy point 
$\varphi=-e/2C$ [see Eq. (9)].
In order to describe the physics at strong orbital magnetic field, i.e., 
away from the degeneracy points and at lower
temperature and more precisely the average grain 
charge $\langle Q\rangle$, we seek to go beyond the effective model in 
Eq. (\ref{heff}).
Indeed, at energy smaller than $T_K[\Delta_z]$, the physics can be 
qualitatively identified
with that of Ref.~[\onlinecite{Gramespacher}]: The Kondo screening of the 
excess 
spin of the small dot by the lead produces an Abrikosov-Suhl resonance at the 
Fermi level, and the small dot plus the lead can be replaced by 
a resonant level with the energy 
$\epsilon\rightarrow 0$ and the resonance width $\sim T_K[\Delta_z]$. Now, one
can still allow for a (weak) residual tunneling matrix element $\hat{t}$
between the grain and the effective resonant level 
(which may be of the same order
as the bare tunneling matrix element $t$ between the small dot and the grain
but its value is difficult to determine accurately).
For an illustration, see Figure 8. Reformulating results of 
Ref.~[\onlinecite{Gramespacher}] for our case and including that 
$T_K[\Delta_z]\ll U_1,U_{-1}$ 
for $N=CV_g/e\ll 1/2$ $(\varphi\ll -e/2C)$, at zero 
temperature we find
\begin{eqnarray}\label{chargef}
\langle Q\rangle &=& e \frac{\Gamma}{\pi} \left( \frac{1}{U_1}
-\frac{1}{U_{-1}}\right)\\ \nonumber
&=& e\frac{\Gamma}{E_c\pi}\frac{4N}{(1-2N)(1+2N)},
\end{eqnarray}

\begin{figure}[ht]
\centerline{
\epsfig{file=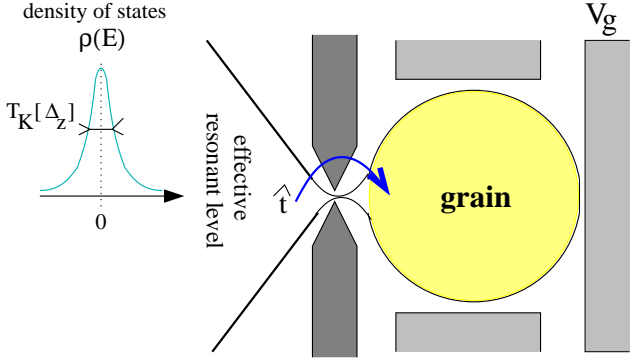,angle=0,height=4.8cm,width=8.3cm}}
\caption{Illustrative view of the effective low-energy model for
almost symmetric barriers away from the degeneracy points: According to
Eq.~(\ref{couplingss}) the charging energy
on the grain inevitably ensures that the spin Kondo coupling $J_0$ between the
bulk lead and the small dot will be the first one to flow off to strong 
couplings at the energy scale $T_K[\Delta_z]$. The grain becomes virtually 
weakly-coupled to an effective {\it resonant level} with a reduced bandwidth 
$\sim T_K[\Delta_z]\ll D$.}
\end{figure}

\hskip -0.3cm with the effective tunneling energy scale 
\begin{equation}
\Gamma=\pi\sum_p \hat{t}^2 \delta(\epsilon_p)
\ll U_1,U_{-1}.
\end{equation}
Since $U_1$ and $U_{-1}$ are of the order of $E_c$ for $N\ll 1/2$, we observe
that the charge smearing far from the degeneracy points is small at low
temperatures. 
Additionally, recall that for $N\ll 1/2$ and zero temperature, at second order 
in $\hat{t}$
the average grain 
charge also
exhibits a (small) linear behavior as a function of $N$ or 
$V_g$ which is slightly distinct from the original Matveev's 
situation (Figure 9).\cite{Matv1,Grabert}

\subsection{Case of asymmetric junctions}

Another interesting perturbation is the explicit symmetry
breaking between the dot-lead and dot-grain tunneling amplitudes. 
To address this issue, it 
is convenient to rewrite
the Kondo Hamiltonian in the most general form
as follows (again $\tau=0$ for
the bulk lead and $\tau=1$ for the grain):
\begin{eqnarray}\label{hef}
H_K &=&\sum\limits_{\tau=0,1}\left( J_{\tau}
\psi_\tau^{\dag}\vec{S}\cdot{\vec \sigma\over 2}\psi_\tau \right)\nonumber\\ 
&+&\sum\limits_{\tau=0,1}\left( \dmi (-1)^\tau V_{z,\tau} T^z 
\psi_\tau^{\dag}\psi_\tau \right)\nn\\ 
&+&{V_{\perp}\over 2}\left[T^+(\psi^{\dag}{\tau^-}\psi)+h.c.\right]\\ \nonumber
&+& \sum\limits_{\tau=0,1}\left( Q_{z,\tau}(-1)^\tau T^z\vec{S}\cdot
(\psi_\tau^{\dag} {\vec\sigma }\psi_\tau)\right)\nonumber\\ 
&+& Q_{\perp}\vec{S}\cdot \left[
T^+(\psi^{\dag}{\tau^-}{\vec{\sigma}}\psi)+h.c.\right].\nn
\end{eqnarray}
The corresponding bare values are embodied by
\begin{eqnarray}
&&J_0=J\ ,\ J_1=K^2J\ ,\ Q_{\perp}=\frac{KJ}{4}\nn\\ &&
V_{\perp}=VK\ ,\ Q_{z,\tau}=V_{z,\tau}=0,
\end{eqnarray}

\begin{figure}[ht]
\centerline{\epsfig{file=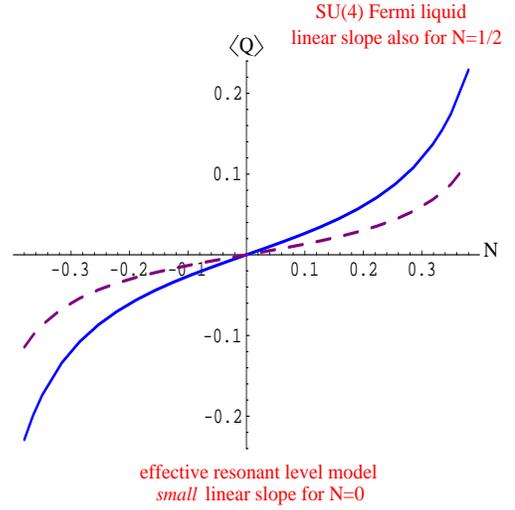,angle=0,height=6.7cm,width=6.6cm}
}
\caption{Profile of the average charge $\langle Q\rangle$ 
on the grain versus $N=CV_g/e$ for (almost) symmetric junctions and 
$T<T_K[\Delta_z]$. 
Again, the SU(4) Kondo 
entanglement between spin and orbital degrees of freedom, e.g.,
at the degeneracy point $N=1/2$ 
produces a Fermi-liquid state and 
the Coulomb staircase exhibits a conspicuous smearing. Away
from the degeneracy points, the physics becomes similar to that of
a resonant level 
weakly coupled to a grain which also ensures a linear (but small) behavior for 
$\langle Q\rangle[N]$ when $N\rightarrow 0$; The full line
curve corresponds to $\Gamma/E_c=0.15$ and the dashed line curve to 
$\Gamma/E_c=0.1$.}
\end{figure}

\hskip -0.3cm where we have introduced the asymmetry parameter 
$K=t_1/t$; $t=t_0$ $(t_1)$ denotes the hopping amplitude between the 
lead (grain) and the small dot. 
Since the asymmetry stands for a marginal perturbation in the RG sense 
it is natural to argue that the SU(4) correlated
ground state is still robust for weak asymmetry between the tunnel junctions.
But, to obtain more quantitative results we yet resort to NRG (Figure 10).
By taking
$V_{\perp}=0.1J$, we can observe that the mixing of spin and orbital degrees
of freedom may 
survive until $K\approx 0.95$; This guarantees an anisotropy of roughly
10$\%$ between the conductances at the tunnel junctions to preserve the SU(4)
fixed point. 
Mostly, the magnetic moment
$\vec{S}$ and the isospin $\vec{T}$ are simultaneously quenched and again the 
spectrum can be understood as a sum of four independent chiral fermion
spectra with phase shift $\pi/4$.

Let us now discuss the case of a quite {\it strong} asymmetry 
between the tunnel junctions. For completeness, we also 
provide the RG equations at second order for this generalized situation

\begin{figure}[ht]
\epsfig{file=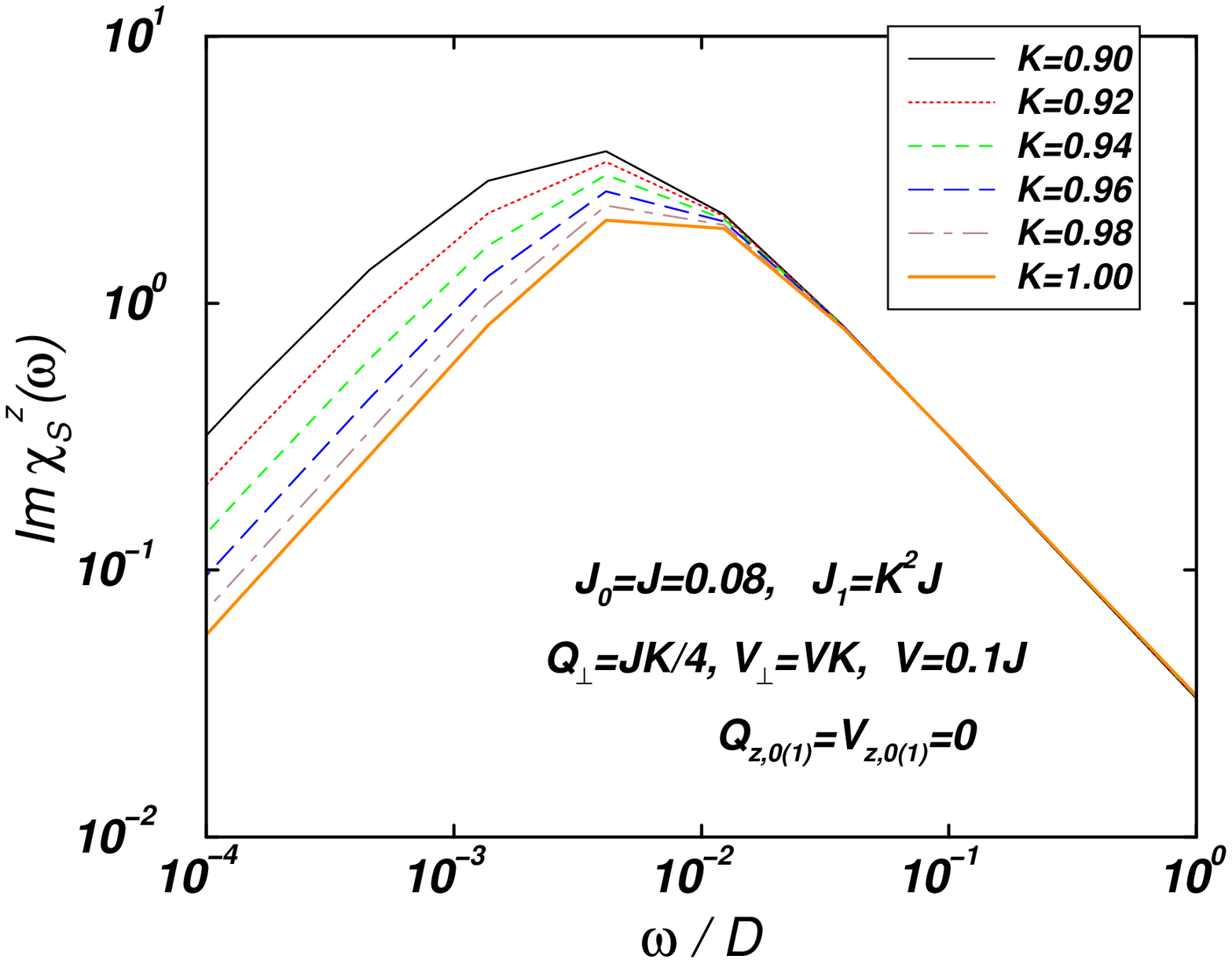,angle=0,height=7cm,width=8.3cm}
\epsfig{file=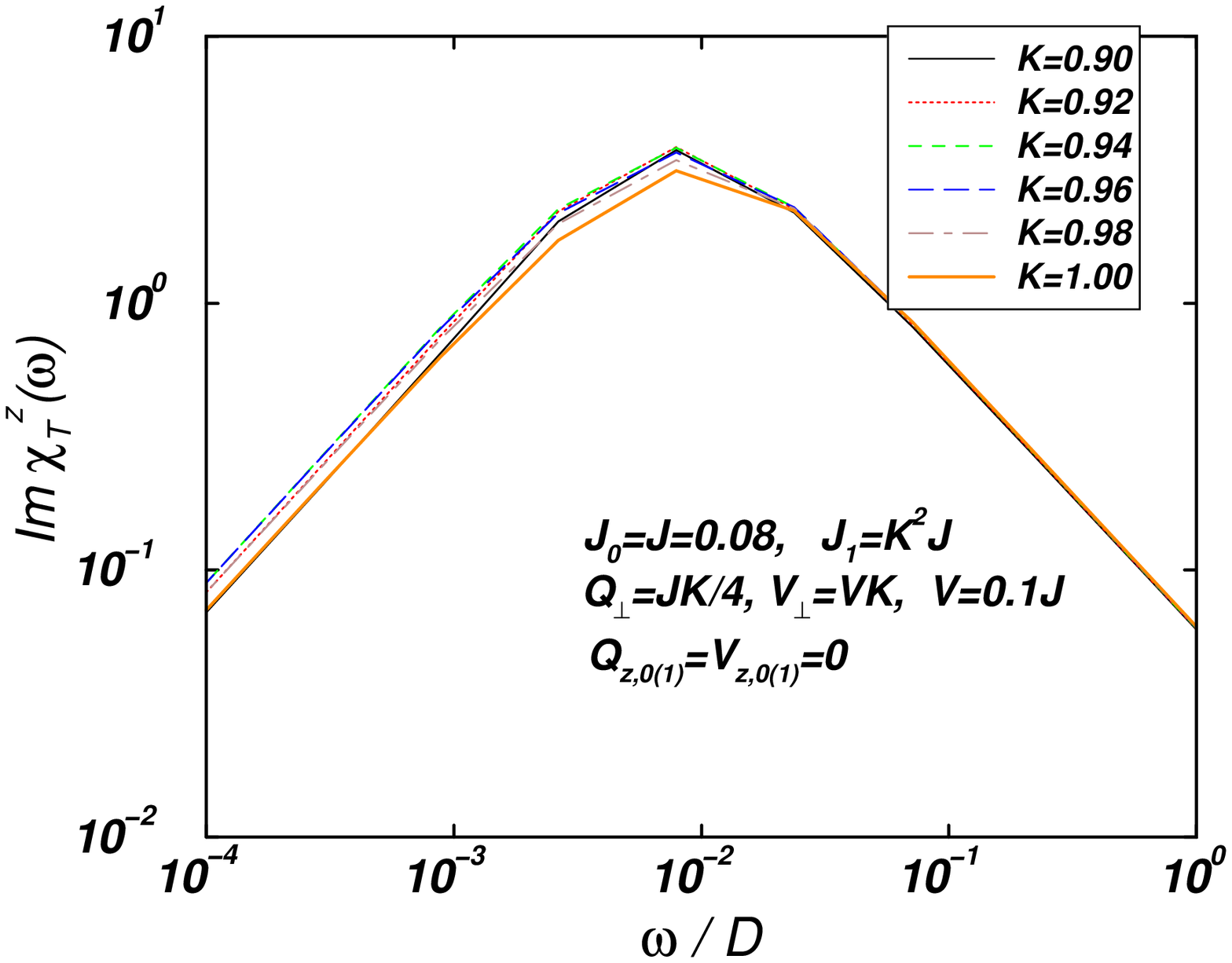,angle=0,height=7cm,width=8.3cm}
\caption{Magnetic and orbital susceptibilities versus $\omega/D$ for 
 close to unity values
of the asymmetry parameter $K=t/t_1$ between the two tunnel junctions. 
The SU(4) ground state is 
stable against the inclusion
of a weak asymmetry between tunnel junctions.}
\end{figure}

\begin{eqnarray}
\frac{d J_\tau}{dl} &=& 
{J_{\tau}}^2 + (Q_{z,\tau})^2 +2{Q_{\perp}}^2 \nonumber \\
\frac{d V_{z,\tau}}{dl} &=& {V_{\perp}}^2 +3 {Q_{\perp}}^2  \nonumber \\
\frac{d V_{\perp}}{dl} &=& \dmi V_{\perp}\left(V_{z,0}+V_{z,1}\right)
+{3\over 2} Q_{\perp}\left(Q_{z,0}+Q_{z,1}\right)\nn\\ 
\frac{d Q_{z,\tau}}{dl} &=& 2J_\tau Q_{z,\tau} +2V_{\perp}Q_{\perp}  
\\
\frac{d Q_{\perp}}{dl} &=& Q_{\perp}\left(J_0+J_1\right) + \dmi Q_{\perp}
\left(V_{z,0}+V_{z,1}\right)\nn\\&+& 
\dmi V_{\perp}\left(Q_{z,0}+Q_{z,1}\right).\nonumber
\end{eqnarray}
At second order, note the equality $V_{z,0}(l)=V_{z,1}(l)=V_z(l)$ 
regardless of the parameter $K$.
Primarily, it is immediate to observe that for $K=1$, we 
recover the previous SU(4) Fermi-liquid flow. 
Now we greatly diminish the
tunneling amplitude between the grain and the small dot, i.e., $t_1\ll t$
($t$ being fixed)
and $K\ll 1$. With the present notations, it is 
clearly transparent that the spin Kondo coupling $J_0=J$ between the bulk 
lead and 
the small dot will be the largest one through the RG flow and becomes of 
order unity at the temperature $T_K[K\ll 1]\sim
D e^{-1/J}=T_K[\Delta_z]$ whereas 
{\it all the other couplings} are still negligible 
that breaks the SU(4) symmetry explicity.

\begin{figure}[ht]
\centerline{
\epsfig{file=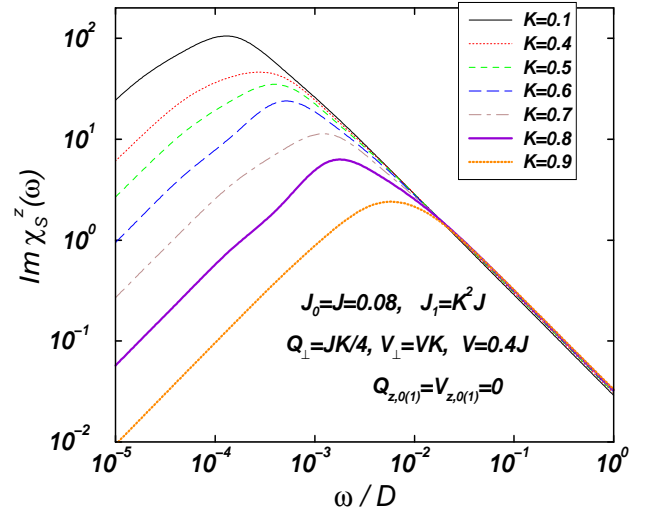,angle=0,height=7.3cm,width=8.3cm}}
\caption{Spin susceptibility versus $\omega/D$ upon increasing the
asymmetry between tunneling amplitudes at the two junctions.}
\end{figure}

It is worth noting at this stage 
that the role of the asymmetry parameter $K$ seems to be 
practically equivalent
to renormalize the orbital splitting $\Delta_z$ (compare Figures 6 and 11). 
The main difference however is that at the degeneracy points
of the grain, one can expect a 
second-stage quenching of the isospin $\vec{T}$ at some lower temperature, 
but obviously this (very) low-temperature regime lies much beyond the range 
of validity of the effective Hamiltonian (\ref{hef}). Furthermore, one can
clearly notice that the previous perturbative result of Eq. (\ref{chargef}) 
diverges if one of the charging energy $U_1$ or $U_{-1}$ approaches zero, i.e., is not applicable. 

In fact, as already noted in Ref.~[\onlinecite{Gramespacher}] it is a very 
difficult task
to find the exact shape of the step of the staircase in the present 
situation of
a grain at a degeneracy point coupled to an effective {\it resonant level}. 
But qualitatively, 
one might expect\cite{Gramespacher} that the physics and the resulting
(two-channel) Kondo energy scale 
should not be so different as those of a grain coupled
to a normal lead with a reduced bandwidth $T_K[K\ll 1]$, via a
hopping matrix element $\hat{t}\sim t_1$:
\begin{equation}
T_K^{2ch}=T_K[K\ll 1] e^{-\gamma/t_1};
\end{equation}
Here $\gamma$ is a constant parameter of the order of unity.
A similar discussion should hold in the opposite regime $K\gg 1$ where
one expect this time the Kondo coupling $J_1$ to first flow
to strong coupling (since it is proportional to $K^2$) at the temperature scale
$T_K[\D_z]\sim De^{-1/J_1}$. Since the conductance between the grain and the lead is still very small at the intermediate energy scale due to the anisotropy,
a second stage quenching of the orbital pseudo spin is expected at a lower energy scale in a similar manner as the case $K\ll 1$.
Unfortunately
for asymmetric junctions, it is difficult to formulate more quantitative
results at low temperatures. A complete renormalization group
calculation starting with the bare Hamiltonian (\ref{anderson}) would be 
necessary. This goes beyond the present analysis.
 Finally, let us mention that for more moderate 
values of $U$ and $\epsilon$, i.e. rather in the resonant level regime of the
small dot, the NRG 
results of Lebanon {\it et al.}\cite{Eran}
still support a two-channel Kondo crossover and the ovserscreening of the
isospin moment in the case of asymmetric junctions.   

\subsection{Large junctions}

We predict that the SU(4) symmetry should 
be still robust for wider junctions characterized by $n>1$ 
transverse channels with almost equal transmission amplitudes, however 
the associated Fermi-liquid
typical energy scale decreases  exponentially with the number of 
conducting modes. For instance, extending results of 
Ref.[\onlinecite{Zarand3}] for
our geometry, we can clearly assess that there will be a {\it unique} 
`effective tunneling mode' in the 
lead (it is some combination of the original tunneling modes in the lead) 
and another {\it unique} `tunneling mode in the box' (also a linear combination
combination of the original modes in the grain): The T=0 effective Hamiltonian
of the model at the degeneracy points of the metallic dot corresponds to 
tunneling between these two modes only with or without spin flip of the excess
spin of the small dot, and all the other modes can be 
neglected. This entirely justifies the emergence of an SU(4) fixed point
at very low temperatures even if the number of modes in the lead or in the
grain is larger than one. However, the ultraviolet cutoff $D$ at which the
effective tunneling mode prevails, must be properly rescaled to\cite{Zarand3}
\begin{equation}
T^{*}[n]=De^{-\alpha n},
\end{equation}
where $\alpha$ is of the order of unity. Unfortunately, this implies that
an SU(4) Kondo singlet can only occur at the much reduced Kondo 
temperature scale 
\begin{equation}
T_K^{SU(4)}[n]\approx T^{*}[n]e^{-1/4J}. 
\end{equation}
Experimentally, in order to maximize chances for observing 
the SU(4) Fermi liquid realm, it
is then more advantageous to consider tunneling junctions with 
one clearly dominant conducting transverse mode.

\section{Discussions and Conclusions}

We have determined exactly the shape of the steps of the Coulomb
staircase for a grain coupled to a bulk lead through a small quantum dot
in the Kondo regime. First, we mapped the problem onto a related
model of two capacitively coupled small quantum dots.\cite{Zarand}
Then, combining both NRG calculations with perturbative scaling approaches
we have shed light on the 
possibility of a stable SU(4) Fermi liquid fixed point occurring at the
degeneracy points of the grain, where
a Kondo effect appears

\begin{figure}[ht]
\centerline{\epsfig{file=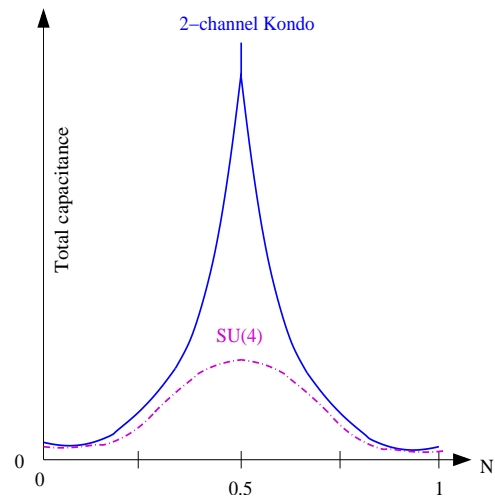,angle=0,height=6.5cm,width=6.4cm}}
\caption{Sketch of the capacitance peaks for our setup with almost symmetric
junctions (dashed line) 
compared to those in the original Matveev's problem (full line).\cite{Matv1}}
\end{figure}

\hskip -0.3cm  simultaneously both in the spin and the 
orbital sectors; This demands symmetric or slightly asymmetric tunnel 
junctions and preferably a single-conducting channel with two 
spin polarizations. 
More generally, as in Ref. [\onlinecite{Zarand}], these results 
bring precursory insight on the realization of Kondo ground states
with SU(N) $(N=4)$ symmetry at the mesoscopic scale.

Let us provide a physical interpretation for the occurrence
of such an SU(4) entanglement. Typically, close to the degeneracy points of
the grain, we have two spin objects, namely the spin $\vec{S}$ of the small dot
and the orbital pseudo-spin $\vec{T}$ of the grain depicting 
the two allowed degenerate charging states. Obviously, when these two spin
objects are uncoupled the symmetry group of the problem is unambiguously 
$SU(2)\otimes SU(2)$. But, as already discussed at length 
in the sequel, in our setting 
spin-flip assisted tunneling events -- i.e., an 
electron from the bulk lead tunnels
onto the metallic grain by flipping the excess spin of the small dot and 
vice-versa -- are very prominent at low energy; This implies that 
the infra-red
fixed point must also reflect a visible spin-orbital mixing. {\it Finally,
it is easy to check that SU(4) is the minimal group allowing spin-orbital
entanglement and which guarantees rotational invariance both in spin and 
orbital spaces}. Our Kondo fixed point
then is rather described
by the quenching
of the hyper-spin $[\vec{S}+{1\over 2}]
[\vec{\hbox{T}}+{1\over 2}]$. 

In a very different context, let us
 mention that SU(4) singlets have also shown up 
in fermion lattice models
where spin and orbital degrees of freedom play a very symmetric 
role.\cite{Li,Azaria}

The major consequence of this enlarged symmetry is
that the ground state is Fermi-liquid like, which considerably smears out the
Coulomb staircase behavior already in the weak tunneling region, 
and in particular, hinders the Matveev logarithmic
singularity\cite{Matv1} to take place (Figure 12). The grain 
capacitance exhibits instead of a logarithmic 
singularity, a strongly reduced peak as a function of the back-gate voltage.
{\it This stands for an irrefutable signature of the formation of
a Fermi-liquid ground state when tunneling through a single-electron box.}
Furthermore, we mightily emphasize that our NRG calculations
markedly reproduce an SU(4) ground state regardless of the particle-hole
asymmetry onto the small dot (Figure 4); More precisely, even in the case of
particle-hole symmetry $2\epsilon+U=0$, the spectrum can be still interpreted
as a sum of four
independent chiral fermions with phase shift $\pi/4$ in agreement with the
SU(4) Fermi liquid theory. This differs from the conclusion of
Ref.~[\onlinecite{Eran}]. 
However, this is not so surprising in the
sense that in their NRG calculations (see, e.g., their Figures 15 and 16),
Lebanon {\it et al.} have studied a rather different limit, 
$U=-2\epsilon$ but $U/E_c\ll 1$, which does
not correspond to our situation of a small dot and a much larger
metallic grain $(U/E_c\gg 1)$. Besides, in the case of symmetric barriers,
they clearly noticed that a moderate
Coulomb repulsion on the small dot already pushes 
the two-channel Kondo regime down to much lower temperature. 

It is also worth to recall that the associated Kondo temperature scale 
$T_K^{SU(4)}$ can be strongly 
enhanced compared to that of the Matveev's original setup which 
maybe ensures the verification of our predictions. 
In particular, for very large U [$U\gg -2\epsilon$ and $V_{\perp}>0$], 
$T_K^{SU(4)}\sim D
\exp-(1/4J)$ may be {\bf larger} than the Kondo scale in the 
conductance experiments across a single small quantum dot\cite{vdw} 
($\sim 1K$), 
and capacitance measurements can be performed much below $100mK$.\cite{Berman}
Additionally, we have checked that the SU(4) Kondo temperature scale 
is considerably reduced for negative values of $V_{\perp}$, i.e., upon
by (moderately) decreasing the on-site interaction $U$ $(U\sim -\epsilon)$,
i.e., by making the small dot larger and larger.\cite{karyn3}
We have carefully discussed the robustness of the SU(4) correlated state
against the inclusion of {\it weak} perturbations like an external magnetic 
field, a deviation from the degeneracy points, or still an
asymmetry in the tunnel junctions. 

\begin{figure}[ht]
\centerline{\epsfig{file=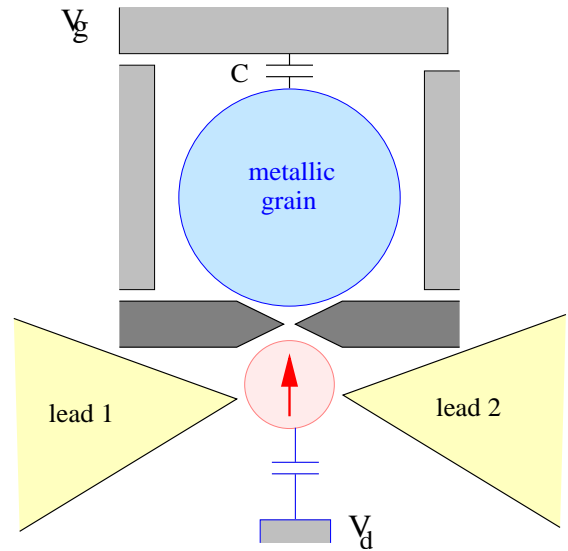,angle=0.0,height=7.3cm,width=
7.4cm}}
\caption{Another mesoscopic double lead setup, candidate for the SU(4) model.
This could be equally performed with vertically coupled dots.\cite{Weiss}}
\end{figure}

Let us now pursue and discuss an interesting crossover. So far, we have 
concentrated
on the situation at and near the degeneracy points of the grain. Let
us now apply a quite strong orbital magnetic field such that we move 
explicitly away from the degeneracy points. Naively, since one suppresses the 
orbital spin-flip terms, one could infer the emergence of a two-channel spin
 Kondo
model through the two Kondo terms $J_0$ and $J_1$; However, in our 
setting with almost symmetric junctions, the 
Schrieffer-Wolff transformation away from the degeneracy points always
ensures $J_0>J_1$; The NRG calculation of Figure 6 clearly reproduces this
expectation. The system then undergoes a one-channel Kondo crossover. 
First, the emergence of
a logarithmic contribution in $\langle Q\rangle$ at quite high temperature
could be potentially observable. Furthermore, 
at low energy, the physics resembles that of a 
resonant level -- induced by the formation of an Abrikosov-Suhl resonance
between the small dot and the 
bulk lead -- weakly-coupled to the grain; We then 
recover a similar situation to 
that of Ref.~[\onlinecite{Gramespacher}].

Another possible realization of our SU(4) model could be still possible in a
multi-lead geometry (Figure 13). Again, this would demand to be at the 
degeneracy points of the grain and to adjust the different tunneling 
junctions. More precisely, following 
Glazman and Raikh,\cite{Raikh} only the even linear
combination of the electron creation and annihilation operators
in the two bulk leads couples to the local site (small dot). The odd linear 
combination can be omitted and conceptually the effective model could be 
rewritten as
in Eq. (\ref{heff}). Let e.g. 
assume that the tunnel junctions between each lead and the small dot
are symmetric. Then, only the linear combination 
$\psi_0=(\psi_{01}+\psi_{02})/\sqrt{2}$ 
will be coupled to the small dot; $\psi_{0i}$ $(i=1,2)$
denotes the electron annihilation operator in each lead. To recover an
SU(4) Kondo fixed point, we infer that the grain-dot tunneling amplitude then
must be approximately $\sqrt{2}$ times that between each lead and the
small dot. This setup is particularly interesting because the
capacitance of the grain and the conductance across the small dot could be
both measured. Furthermore, blocking completely
the opening between the grain
and the small dot, one could recover a more usual Fermi liquid behavior
with SU(2) spin symmetry when
measuring the conductance across the small dot, and 
observe a net reduction
of the Kondo energy scale compared to the SU(4) case due to spin orbital
decoupling. 

Note that this geometry -- away from the degeneracy points of the grain -- 
has been previously discussed by Oreg and
Goldhaber-Gordon as a potential candidate for the appearance
of a two-channel (spin) Kondo regime
in a conductance measurement.\cite{oreg} This requires meticulous fine-tuning
of the gate voltages and tunnel junctions to equalize the coupling
to the two channels (grain plus even linear combination of the leads).

Definitely, the potential observation
of a two-channel Kondo effect in artificial nanostructures would be an 
important issue\cite{Matv1,ralph,achim,kim} 
since the emergent non-Fermi liquid behavior is very intriguing
and so far 
difficult to observe with real magnetic impurities due to the intrinsic
channel anisotropy.\cite{no}  In our setting, another interesting 
experiment to do in order to have potentially 
access to a two-channel (charge) Kondo behavior
would be to stay at the degeneracy points of the grain and then
progressively to shift the impurity level $\epsilon$ on the dot
(which can be tuned via the gate voltage $V_d$ of the small dot)
to the Fermi energy, i.e., to reach the mixed-valence (=resonant level)
limit for the small dot.\cite{Eran}

\begin{widetext}

\acknowledgments

Part of this work was performed during the Quantum Impurity conference
meeting in Dresden (April 2003). K.L.H was supported in part by NSERC and 
acknowledges constructive discussions with K. Matveev.
P.S. acknowledges interesting discussions with P. Brouwer, L. Glazman and P. 
Sharma. L.B. acknowledges the support of `Spintronics' RT Network of the 
EC RTN2-2001-00440 and Hungarian Grant No. OTKA T034243.

\appendix
\section{Perturbative calculations}

Here, we derive explicitly the perturbative result of Eqs.~(\ref{pert}) and
(9). 
We essentially focus on the Kondo term; the perturbation theory for the
direct hopping term $V$ can be found in Ref.~[\onlinecite{Matv1}]. First, it
is accurate to rewrite the Kondo term in real space as: 
\begin{equation}
H_K = \sum_{\alpha\beta}\left[
\sum_{j=0,1} \frac{J_{j}}{2}\vec{S}\psi^{\dagger}_{j\alpha}
\vec{\sigma}_{\alpha\beta}\psi_{j\beta}+\frac{\tilde J_{10}}{2}\vec{S}\left(
\psi^{\dagger}_{0\alpha}\vec{\sigma}_{\alpha\beta}\psi_{1\beta}+h.c.
\right)\right],
\end{equation}
where $\psi_{0\alpha}=\sum_k a_{k\alpha}$ and $
\psi_{1\alpha}=\sum_p a_{p\alpha}$. The granule charge operator reads
$\hat{Q}=e\sum_{\alpha}\psi^{\dagger}_{1\alpha}\psi_{1\alpha}$.
Now, let $|0\rangle$ denote the ground state of the unperturbed Hamiltonian
with $t=-\infty$. The first order correction $|1\rangle$ to $|0\rangle$ then 
reads:\cite{Gramespacher}
\begin{equation}
|1\rangle = -i \int_{-\infty}^0 \ dt H_K(t)|0\rangle,
\end{equation}
$H_K$ being taken in the interaction representation.
The expectation value of the charge on the dot however is second order in the
Kondo coupling; Indeed, we easily get $\langle 0|\hat{Q}|1\rangle =0$. 
Therefore, the most leading contribution takes the form 
$\langle\hat{Q}\rangle_2=\langle 0|\hat{Q}^{(1)}|1\rangle$, where 
$\hat{Q}^{(1)}$ is the first order 
correction to the charge operator on the dot. This can be computed using the 
identification:
\begin{equation}
\hat{Q}^{(1)} =\int_{-\infty}^0 \hat{J}(t)dt \ \hbox{with} \ 
\hat{J}(t)=i[H_K,\hat{Q}].
\end{equation}
$\hat{J}$ must be identified as the effective current operator mediated by
the Kondo coupling. This results in
\begin{equation}
\hat{Q}^{(1)} = ie\frac{\tilde J_{10}}{2}\sum_{\alpha\beta}
\int_{-\infty}^0 dt \left[\vec{S}\psi^{\dagger}_{0\alpha}(t)
\vec{\sigma}_{\alpha\beta}
\psi_{1\beta}(t)-\vec{S}\psi^{\dagger}_{1\alpha}(t)
\vec{\sigma}_{\alpha\beta}\psi_{0\beta}(t)\right].
\end{equation}
The expectation value of the charge on the dot is then to second order in
the coupling to the impurity
\begin{eqnarray}
\langle\hat{Q}\rangle_2 &=&
 e \frac{({\tilde J}_{10})^2}{4}\sum_{a,b}\sum_{\alpha\beta} 
\int_{-\infty}^0 dt_1 \int_{-\infty}^0 dt_2\
\langle S^a(t_1)S^b(t_2)\sigma^a\sigma^b\rangle \hbox{\Large{[}}\langle
\psi^{\dagger}_{0\alpha}(t_2)\psi_{0\alpha}(t_1)\rangle
\langle \psi_{1\alpha}(t_2)\psi^{\dagger}_{1\alpha}(t_1)\rangle
 \\ \nonumber
& & \hskip 7.7cm -\langle \psi_{0\beta}(t_2)\psi^{\dagger}_{0\beta}(t_1)\rangle
\langle \psi^{\dagger}_{1\beta}(t_2)\psi_{1\beta}(t_1)\rangle  \hbox{\Large{]}}
\\ \nonumber
&=& e \frac{3(\tilde J_{10})^2}{8}\int_{-\infty}^0 dt_1 \int_{-\infty}^0 dt_2
\left[\langle
\psi^{\dagger}_{0}(t_2)\psi_{0}(t_1)\rangle
\langle \psi_{1}(t_2)\psi^{\dagger}_{1}(t_1)\rangle
-\langle \psi_{0}(t_2)\psi^{\dagger}_{0}(t_1)\rangle
\langle \psi^{\dagger}_{1}(t_2)\psi_{1}(t_1)\rangle \right]
\end{eqnarray}
where the averages are taken over the ground state of the uncoupled system.
It is advantageous to Fourier transform the problem as:
\begin{equation}
\langle\hat{Q}\rangle_2 =  -e \frac{3({\tilde J}_{10})^2}{8}
\sum_{p,k}\int_{-\infty}^0 dt_1 \int_{-\infty}^0 dt_2
\left[\langle a_{k}(t_2)a_k^{\dagger}(t_1)\rangle\langle 
a_p^{\dagger}(t_2)a_p(t_1)\rangle-
\langle a_{p}(t_2)a_p^{\dagger}(t_1)\rangle\langle 
a_k^{\dagger}(t_2)a_k(t_1)\rangle\right]
\end{equation}
where the momentum indices $p$ and $k$ respectively refer to the grain and
to the reservoir.
Using the Green's functions of the isolated grain:
\begin{eqnarray}
\langle a_{p}^{\dagger}(t_2)a_{p}(t_1) \rangle
&=& \Theta(-\epsilon_p)
e^{i(\epsilon_p-U_{-1})(t_2-t_1)} \\ \nonumber
\langle a_{p}(t_2)a_{p}^{\dagger}(t_1)\rangle &=&\Theta(\epsilon_p)
e^{-i(\epsilon_p+U_{1})(t_2-t_1)},
\end{eqnarray}
where again $U_{1}$ and $U_{-1}$ embody the energies to add an electron and
hole onto the grain, we finally find:
\begin{eqnarray}
\langle\hat{Q}\rangle_2 &=& -e 
\frac{3({\tilde J}_{10})^2}{8}\sum_{k,p}\left[
\frac{\Theta(\epsilon_k)\Theta(-\epsilon_p)}
{(\epsilon_k-\epsilon_p+U_{-1})^2}-\frac{\Theta(-\epsilon_k)\Theta(
\epsilon_p)}{(\epsilon_p-\epsilon_k+U_{1})^2}\right]\\ \nonumber
&=& e \frac{3({\tilde J}_{10})^2}{8}\ln\left({e/2C-
\varphi\over
e/2C+\varphi}\right).
\end{eqnarray}
$\Theta$ is the usual Heavyside function.
Density of states in the grain and in the lead have been
assumed to be equal and taken to be 1 for simplicity.

Now, we briefly want to show that cubic orders involve logarithmic divergences
both associated with the Kondo coupling and with the proximity of a degeneracy
point in the charge sector. More precisely, let us focus on the specific 
contribution in $J_{0}{\tilde J_{10}}^2$ for the term 
$\langle\hat{Q}\rangle_3=\langle 0|\hat{Q}^{(1)}|2\rangle$, with
\begin{eqnarray}
|2\rangle &=& -\frac{1}{2}\int_{-\infty}^0 dt_1\int_{-\infty}^0 dt_2\
T[H_K(t_1)H_K(t_2)]|0\rangle\\ \nonumber
&=& -\frac{J_{0}\tilde J_{10}}{2}\sum_{a,b}\sum_{\alpha\beta}\sum_{\mu,\nu}
\int_{-\infty}^0 dt_1\int_{-\infty}^0 dt_2\
T[S^a(t_1)S^b(t_2)]T[\psi^{\dagger}_{0\alpha}(t_1)
\frac{{\sigma}^a_{\alpha\beta}}{2}\psi_{0\beta}(t_1)
\psi^{\dagger}_{0\mu}(t_2)
\frac{{\sigma}^b_{\mu\nu}}{2}\psi_{1\nu}(t_2)]|0\rangle\\ \nonumber
&=&  -\frac{J_{0}\tilde J_{10}}{2}\sum_{a,b}\sum_{\alpha\beta}\sum_{\mu,\nu}
\int_{-\infty}^0 dt_1\int_{-\infty}^0 dt_2\
T[S^a(t_1)S^b(t_2)][T\langle \psi^{\dagger}_{0\alpha}(t_1)\psi_{0\nu}(t_2)
\rangle \delta_{\alpha\nu}
\psi_{0\beta}(t_1)\psi^{\dagger}_{1\mu}(t_2)\frac{{\sigma}^b_{\mu\nu}}{2}
\frac{{\sigma}^a_{\nu\beta}}{2}]|0\rangle\\ \nonumber
&=& +\frac{J_{0}\tilde J_{10}}{2}\sum_{c}\sum_{\alpha}\sum_{\mu,\beta}
\int_{-\infty}^0 dt_1\int_{-\infty}^0 dt_2\ S^c sgn(t_1-t_2)
T\langle \psi^{\dagger}_{0\alpha}(t_1)\psi_{0\alpha}(t_2)\rangle
\psi^{\dagger}_{1\mu}(t_2)\frac{\sigma^c_{\mu\beta}}{2}
\psi_{0\beta}(t_1)|0\rangle \\ \nonumber
&\approx&  -i{J_{0}\tilde J_{10}}\sum_{c}\sum_{\mu,\beta}
\ln\left(\frac{D}{k_B T}\right)\int dt_1\ S^c
\psi^{\dagger}_{1\mu}(t_1)\frac{\sigma^c_{\mu\beta}}{2}
\psi_{0\beta}(t_1)|0\rangle.
\end{eqnarray}
It becomes then obvious that $|2\rangle$ is (almost) 
proportional to $|1\rangle$; 
It is straightforward to show that this induces a third-order 
correction for the charge on the grain 
\begin{equation}
\langle\hat{Q}(T)\rangle_3 \propto J_{0}(\tilde J_{10})^2\ln
\left(\frac{D}{T}\right) 
\ln\left({e/2C-
\varphi\over e/2C+\varphi}\right).
\end{equation}
Note that the appearance of the extra $\ln\left(D/T\right)$ factor
clearly 
stems from the prominent renormalization of the lead-dot spin Kondo coupling
$J_0$ on a charge plateau. 

\end{widetext}

\end{document}